# Combining spectral analysis and narrow band pass filtering of sunspot number records to forecast the next solar grand minimum.

**Ian Edmonds[1] and Peter Killen[2]**

(1) Queensland University of Technology, Brisbane, Australia (retired), ian13edmonds@gmail.com

(2) University of the Sunshine Coast, Sunshine Coast, Australia (retired)

## Abstract

The paper develops a method for forecasting grand minima in sunspot number (SSN). The method is applied to the annual SILSO SSN record, 1700 - 2024, to obtain a high resolution forecast and to a reconstructed annual SSN record, 971 - 1899, to obtain a low resolution forecast. Fourier analysis identifies eleven near-decadal periodicities, in the range from 8 to 15 years, in the SILSO record and four centennial scale periodicities, between 60 and 800 years, in the reconstructed SSN record. Components at each period are isolated by a new method of narrow band pass filtering. The components are fitted with sinusoids for back/forward projection. Forward projection from each record forecasts a grand solar minimum from 2030 to 2150. Short term features within the grand minimum are less certain. The eleven periodicities in the SILSO record are separated into octal, decadal and decadal-plus groups. The four components in the decadal group are, on average, higher in amplitude by a factor of three than components in the other two groups. During grand maxima the decadal group components constructively interfere and dominate SSN variation. Following grand maxima the four decadal group components move into destructive interference, a grand minimum in SSN occurs and the summed amplitudes of the octal and decadal-plus groups can match or exceed the summed decadal contribution. Short term change in SSN variation, specifically the Waldmeier Effect, is explained in terms of interference between the contributions of the octal and decadal groups.

## 1. Introduction

The number of sunspots on the surface of the Sun has been used as the primary measure of solar activity since the invention of the telescope. The temporal change in SSN is characterized by a cycle with a period of approximately 11 years (the Schwabe cycle) which marks the time between intervals of high solar activity, (Clette et al 2014, Hathaway 2015). The Schwabe cycle varies in amplitude and cycle length. Extended intervals of weak solar cycles (low SSN), grand minima, occur between intervals of strong solar cycles, grand maxima. The occurrence and duration of grand minima/maxima is complex, (Usoskin et al 2007, Inceoglu et al 2015, Usoskin et al 2016, Inceoglu 2024), and is variously described as stochastic, chaotic or deterministic or some mixture of all three, and is poorly understood (Stefani et al. 2024). Due to the effects of space radiation on personnel and equipment, the possibility of forecasting Schwabe cycles and grand minima/maxima is of interest. Numerous methods to predict the form of the next Schwabe cycle have been developed, most relying on some aspect of the most recent cycle, a precursor, to predict amplitude and occurrence of the next cycle, (Clilverd et al 2006, de Jager and Duhau 2009, 2012, Cameron and Schussler 2008b, Hathaway and Upton 2016, Pesnell 2018, Kitiashvili 2020, Bisoi et al 2020, Svalgaard 2020, Courtillot et al 2021, Nandy 2021, Jain et al 2022, Mursula 2023, Asikainen and Mantere 2023, Cao et al 2024, Rodriguez et al 2024a &b). The current view is that "solar activity is neither stationary, nor multi-periodic. It contains a significant stochastic component, which makes the SN series non-stationary and limits the prediction horizon to about one cycle", (N. Krivova, personal communication).

Forecasts of Schwabe cycle amplitude based on different methods produce disparate results that vary (Pesnell 2016, 2018, Foxon 2025), by as much as a 5:1 ratio, according to Nandy (2021), who also suggests that "*reasonably accurate predictions are possible only for the next sunspot cycle, and not beyond*", and Karak (2024) who concludes "*a reliable prediction of the solar cycle amplitude beyond one cycle is impossible*". Evidently the science of high resolution forecasting of solar cycles is still at an early stage. However, there have been a number of low resolution forecasts for as far ahead as 500 years (e.g. Steinhilber and Beer 2013, Reikard 2020).

Here we propose a new method of prediction based on Fourier analysis (FFT) and very narrow band pass filtering to obtain the components of SSN at the periods identified by FFT. On the assumption that the periodicities identified by FFT are persistent, the eleven components are fitted with sinusoids for projection beyond the observed record with destructive/constructive interference between the sinusoids generating future grand minima/maxima in SSN. We apply this method to two records of SSN variation, (1), the SILSO annual average SSN record, 1700 to 2024, where the method utilizes the near-decadal range periodicities and (2), the reconstructed annual average SSN record 971 to 1899 (Usoskin et al. 2021), where the method utilizes centennial-range periodicity. The application of the method to both records forecasts an extended grand minimum in SSN during the current century. To provide a simple model of grand minima formation we show, in the case of the near-decadal periodicities, that several components equally spaced in the frequency domain transform into the time domain as a variation with features similar to the features of grand minima/maxima variation in reconstructed records of solar activity. The eleven near-decadal periodicities in SILSO SSN can be separated into octal, decadal and decadal-plus groups and we show that interference between the decadal and the octal components reproduces the individual Schwabe cycle characteristics known as the Waldmeier Effects, (Waldmeier 1935, Takalo and Mursula 2018).

Section 2 outlines data sources and methods. Section 3 presents the spectral content of SILSO SSN and reconstructed SSN, the octal, decadal and decadal-plus and centennial range components, the generation of Waldmeier Effects, and the projection method leading to the forecasts of the next grand minimum. Section 4 is a discussion that includes models of grand minima formation and the Waldmeier Effect. Section 5 is a summary and conclusion.

## 2. Data and Methods

The SILSO SSN annual record, 1700.5-2024.5, was downloaded from https://www.sidc.be/SILSO/datafiles . Spectral analysis was obtained with the FFT implemented in the DPlot application. Digital notch filtering, (Press et al 2007), was also implemented in the DPlot application. The procedure to obtain a band pass filtered version of SSN involved: (1) The mean SSN was subtracted from each SSN record to obtain the anomaly from the mean. (2) A Press notch filter was applied to the resulting anomaly with the bandwidth of the notch filter specified as a percentage of the frequency of the periodicity assessed by FFT. (3) The output of the notch filter was subtracted from the SSN anomaly to provide a band pass filtered version of SSN, referred to here as the inverse notch filtered (INF) version. For decadal-range periodicity, narrow band INF was obtained with the band pass of the notch filter set at 2% of the centre frequency for each of the four periodicities in the decadal range. For example, to obtain the narrow band component of SILSO SSN at 11.0 years the Press notch filter in the Dplot application was set to a centre frequency 0.0909 $yr^{-1}$ and the bandwidth of the notch filter to 0.0018 $yr^{-1}$. Wavelet analysis was obtained with the application WaveletComp, (Roesch and Schmidbauer, 2018), to the SILSO reconstructed SSN.

**3. Results.** Sections 3.1 specifies the near-decadal range periodicities and corresponding INF components in the SILSO SSN record. Section 3.2 fits sinusoids to the components and projects these forward to obtain a high resolution forecast of SSN between 2024 and 2200. Section 3.3 establishes the relation of decadal-range components to the various Waldmeier Effects. Section 3.4 applies the method of component identification and sinusoid fitting to a reconstructed SSN record to obtain a low resolution forecast of SSN to 2200.

**3.1 Characterising the spectral content of SILSO SSN.** The SSN anomaly is shown as the full black line in Figure 1A. The periodogram of the SSN anomaly in the near-decadal range of interest here is shown (black full curve) in Figure 1B. For this paper the periodicities evident in Figure 1B were divided into three groups, an octal group, a decadal group and a decadal-plus group, as shown in Table 1.

Rozelot (1993) found essentially the same decadal group periodicities in SSN (i.e., FFT peaks at 10.01, 10.56, 11.10 and 12.05 years). Similarly, Scafetta (2012) and Scafetta and Bianchini (2023, 2026) identify the same four decadal group periodicities in the SILSO SSN record extending from 1700. The wavelet analysis of the SSN anomaly, Figure 1C, shows how the periodicity within the range of each of the three groups varies with time. While the periodicity is primarily decadal it occasionally swings to the octal range, for example around 1780, or to the decadal-plus range, for example around 1800.

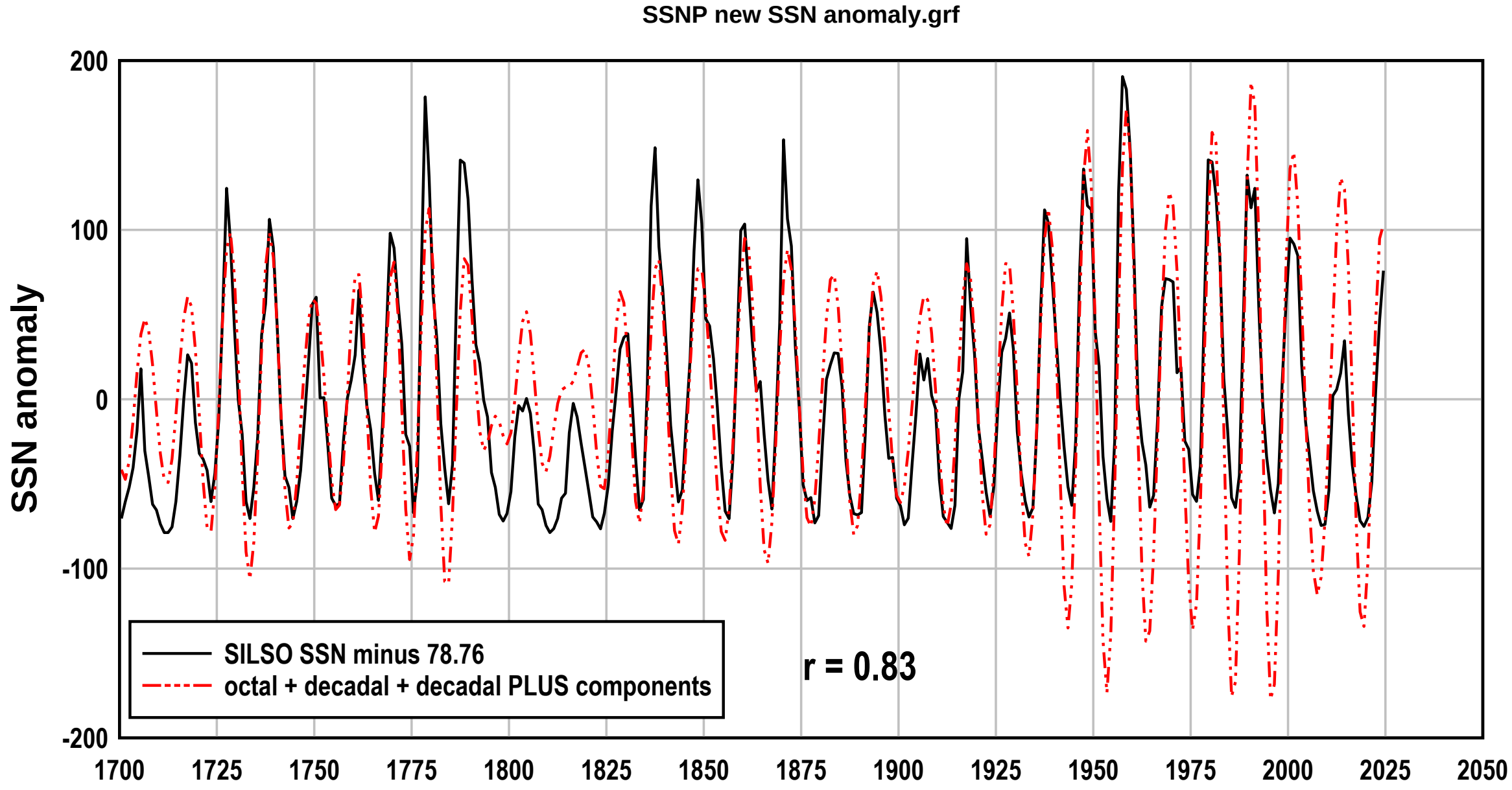

(A)

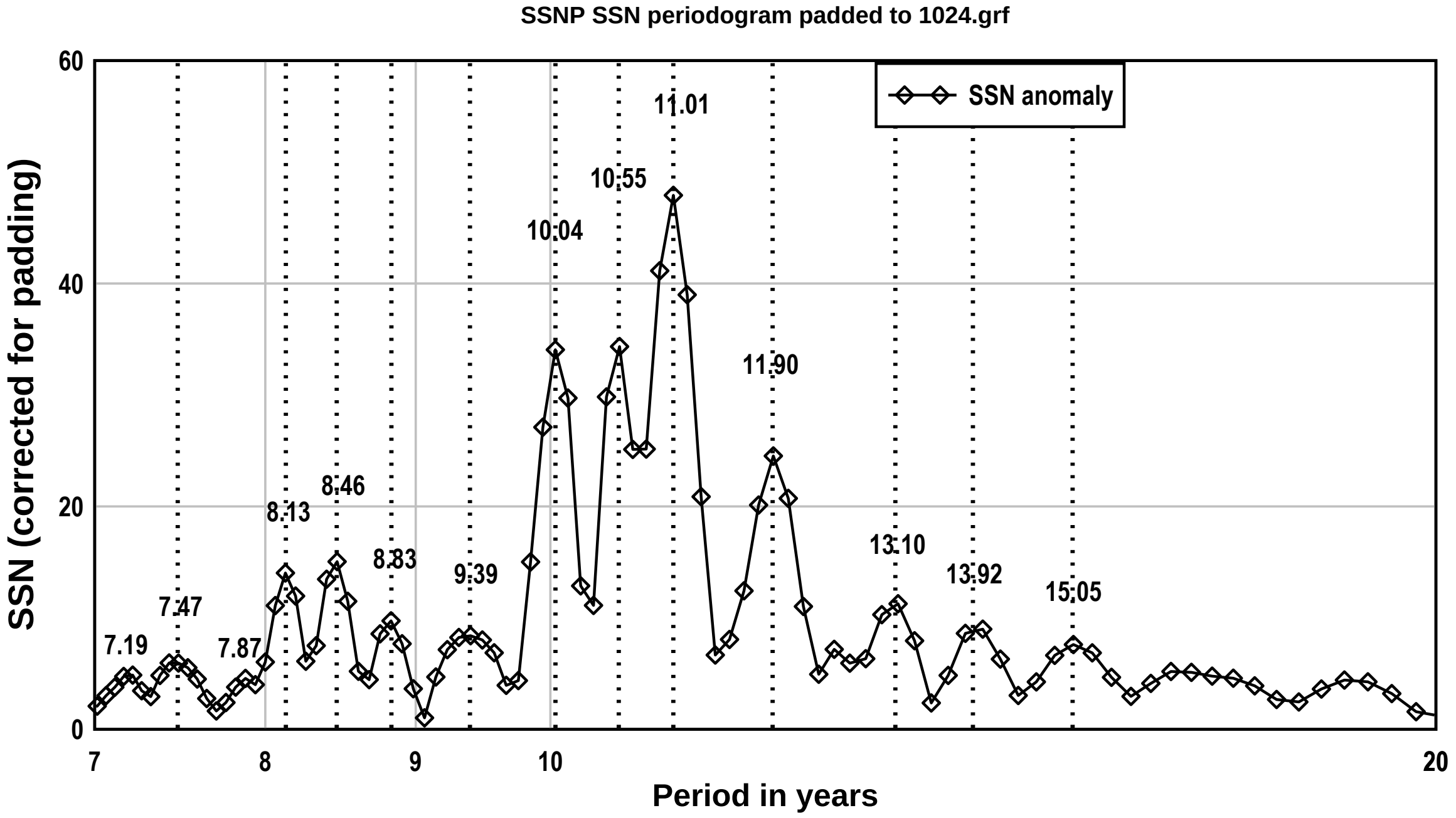

**(B)**

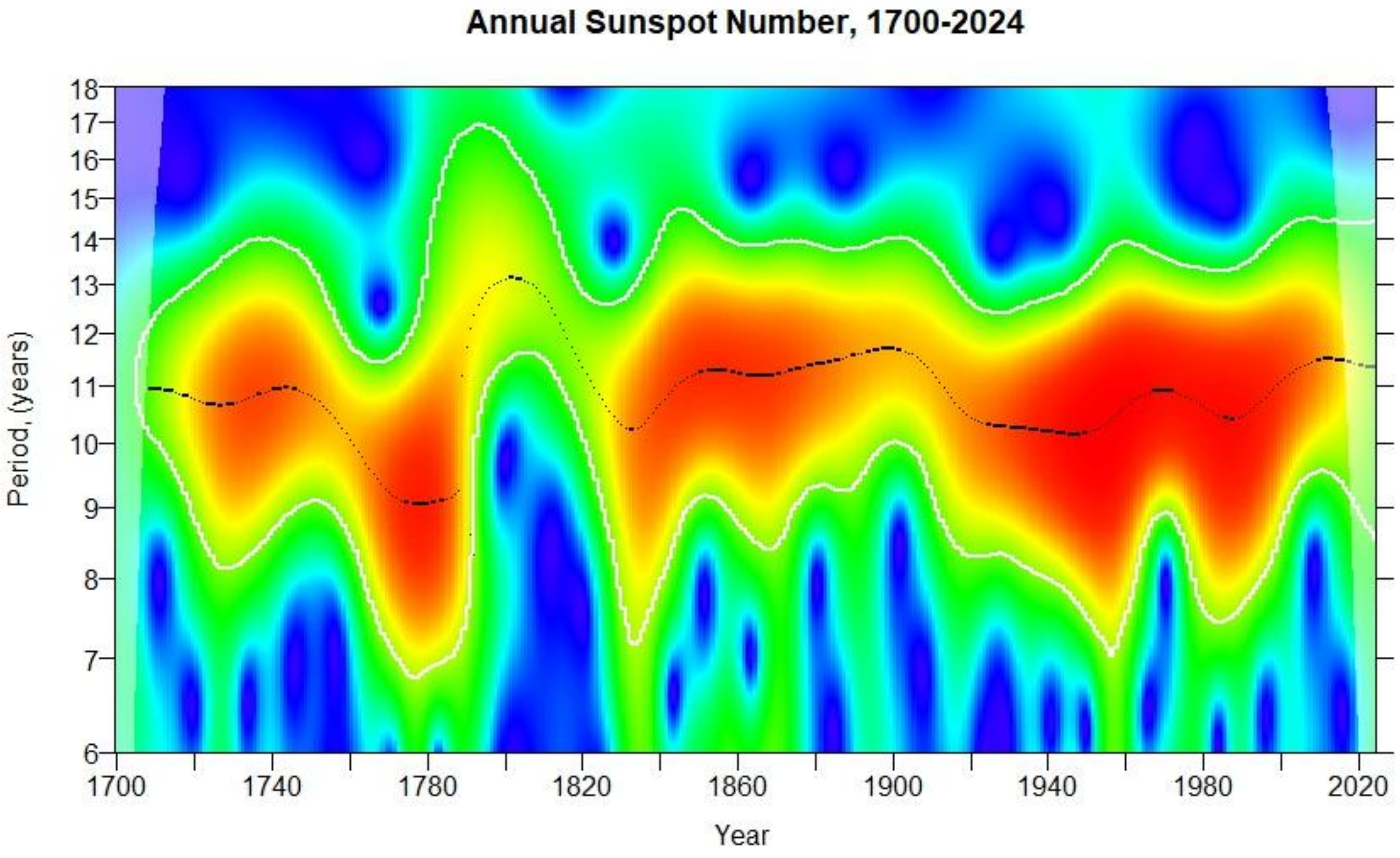

(C)

**Figure 1. (A) Comparisons of the time variation of SILSO SSN (less the mean, i.e. the SSN anomaly) and the sum of the four decadal, four octal and three decadal-plus components. Notable features are the Dalton Minimum around 1810 and the Modern Grand Maximum around 1980. (B) The periodogram of SILSO SSN. While peaks in the 7 – 8 year period range are labeled these are not included within the octal group. (C) The wavelet analysis of SILSO SSN shows that the spectral power is mainly in the 10 - 12 year decadal period range but occasionally, as indicated by the power ridge (black line), power shifts strongly to the octal period range, for example, the interval around 1780, or to the decadal-plus range, the interval around 1800.**

Narrow band INF filtering was used to obtain components at the periodicities labelled in Figure 1B. The components are shown in Figures 2A, 2B and 2C as well as the sum of the components in each group. The variations of the group sums are consistent with the wavelet analysis, (Figure 1C). For example, the octal components, Figure 2B, constructively interfere at around 1780 and the decadal-plus components, Figure 2C, constructively interfere at around 1800 resulting in the swings in periodicity with time evident in the wavelet analysis.

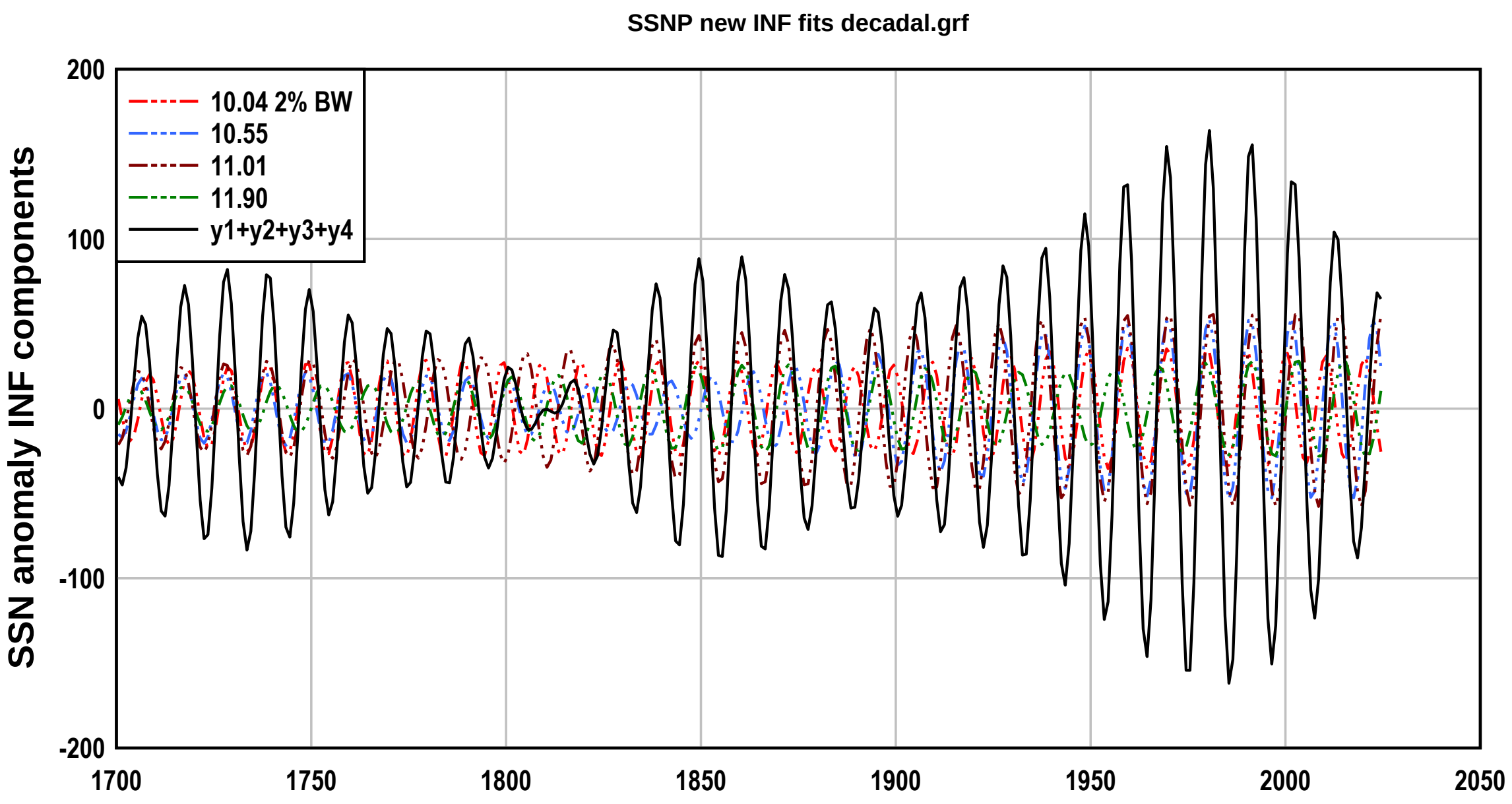
SSNP new INF fits decadal.grf
10.04 2% BW
10.55
11.01
11.90
y1+y2+y3+y4
SSN anomaly INF components
200
100
0
-100
-200
1700
1750
1800
1850
1900
1950
2000
2050

(A)

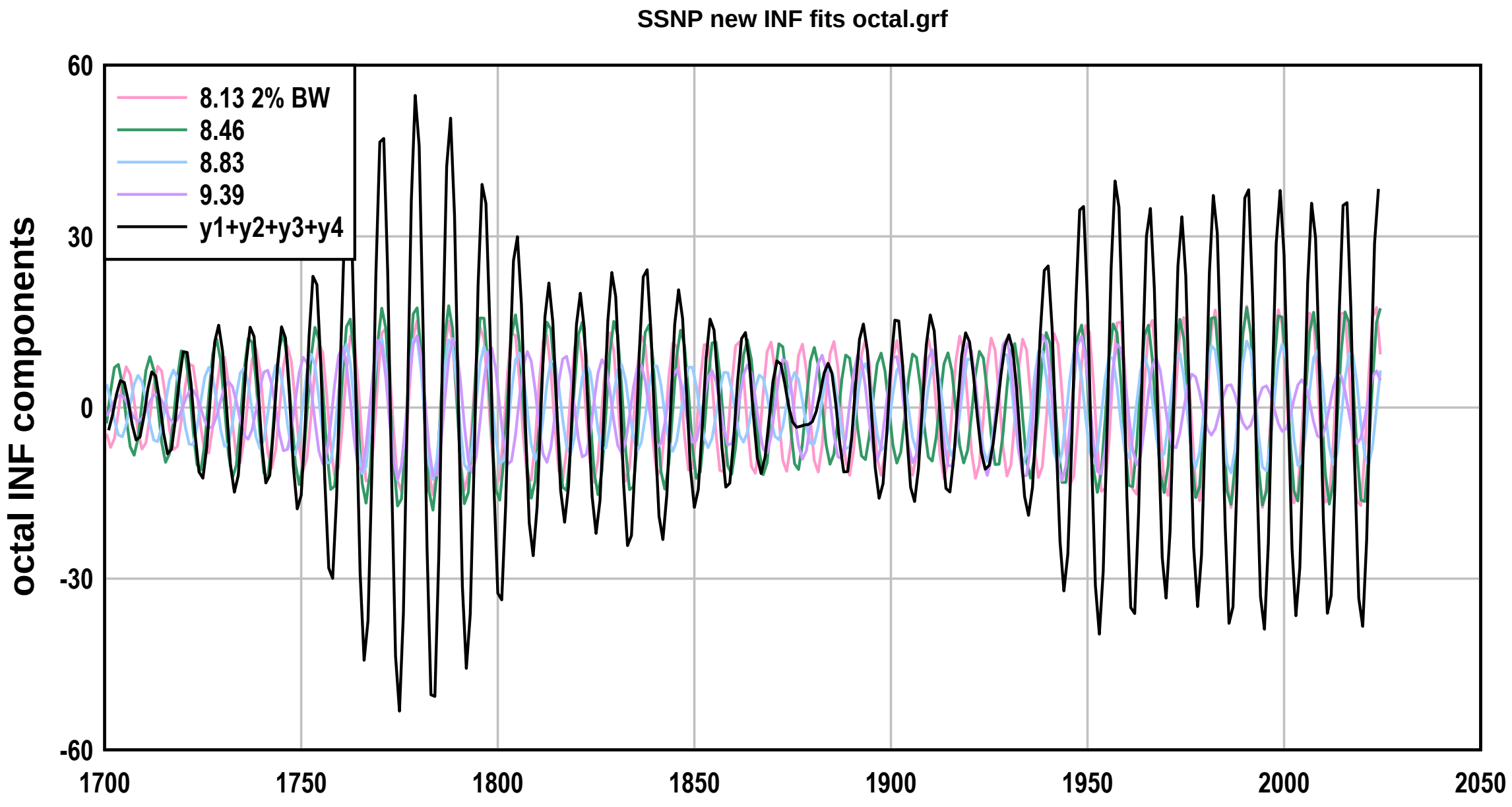
SSNP new INF fits octal.grf
8.13 2% BW
8.46
8.83
9.39
y1+y2+y3+y4
octal INF components
60
30
0
-30
-60
1700
1750
1800
1850
1900
1950
2000
2050

(B)

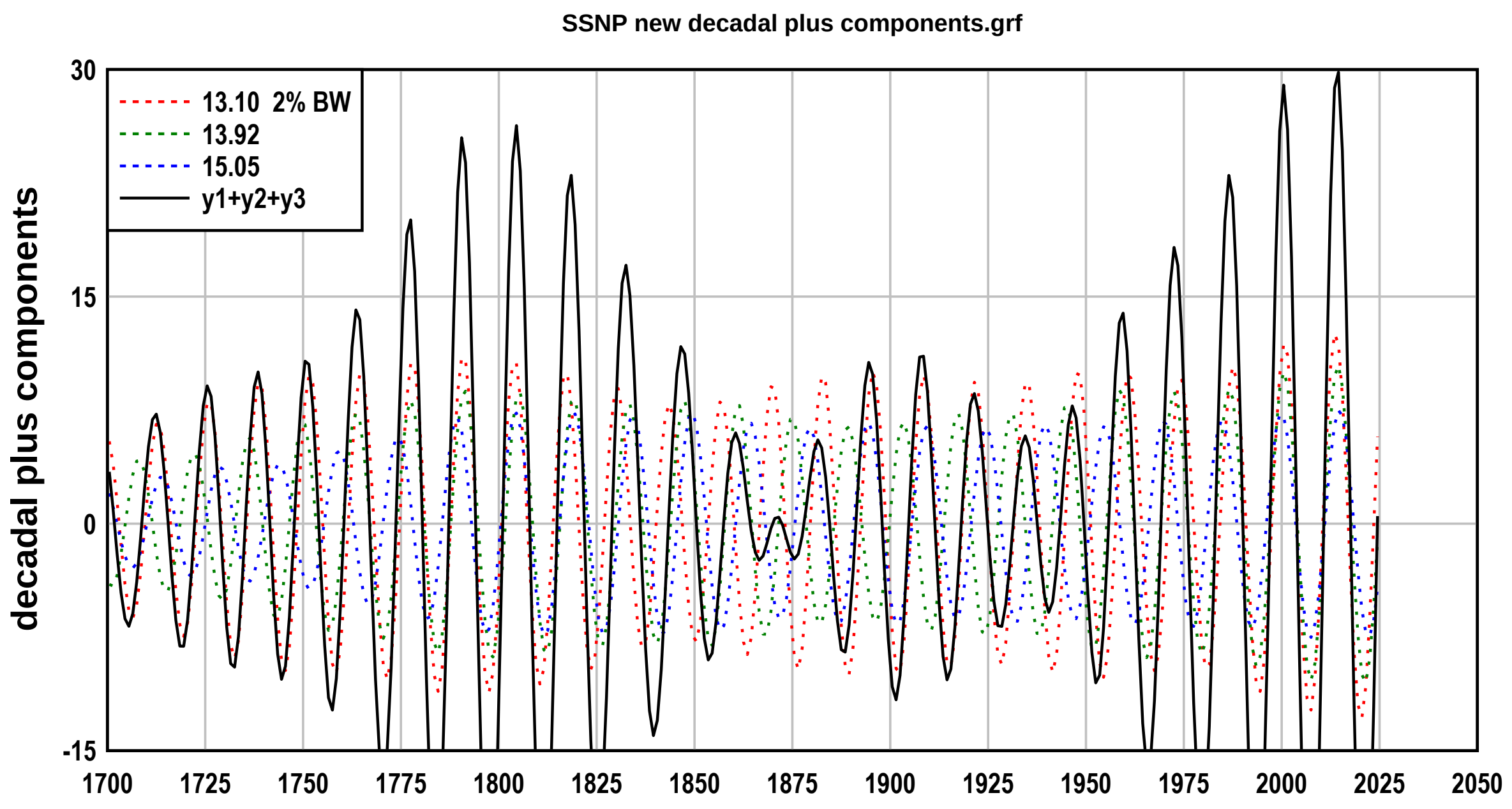

(C)

**Figure 2. A, B and C show the components obtained by the INF method at 2% band width in the decadal, octal and decadal-plus groups respectively. The sum of the component in each group is shown by the full black curves. Note that the octal and decadal-plus components occasionally become dominant (e.g. the octal group between 1750 and 1800 and the decadal-plus group around 1800).**

The sum of all eleven INF components is compared with the SSN anomaly in Figure 1A, (broken red line and solid black line respectively). The correlation coefficient, r = 0.83, indicates that the component combination closely replicates the SSN anomaly. The components were obtained by the INF method of band pass filtering with the pass band set at 2% of centre frequency of each periodicity. The amplitude of an INF component depends on how closely the filter pass band matches the half width of the selected periodicity. The half width varies for each peak in Figure 1B but is approximately 2% of the centre frequency. Ideally, the amplitude of each component would be adjusted so that the periodogram of the component sum matches in amplitude the periodogram of the SSN anomaly. The adjustments required in the present case are small, octal, x 0.89, decadal, x 0.72, and decadal-plus, x 0.80. When the group adjustments are made, the result is the match shown by the broken red curve in Figure 1B. However, the correlation between the component sum and the SSN anomaly is unchanged by the adjustment (i.e. the correlation remained at r = 0.83). Similarly, the forecast of a grand minimum obtained below was not significantly changed by the adjustment of amplitude for filter bandwidth, mainly because the four decadal components destructively interfere during grand minima. It appears that the interference between components depends mainly on the phase and not on the amplitude of the components. For this reason, the components as obtained by the 2% band width INF method (Figures 2 A, B and C) are used in the analysis that follows. By contrast, in section 3.4 where a low resolution forecast is made using centennial scale periodicities obtained from a long reconstructed SSN it becomes necessary to adjust the amplitude of the INF components to match the amplitude of peaks in the SSN periodogram.

Table 1. Parameters of sinusoids fitted to the components at solar cycle 24.

| Group | Frequency ($year^{-1}$) +/- 0.0005 | Period (years) | Phase (year) +/- 0.1 | Amplitude (SSN) +/- 1 |
|---|---|---|---|---|
| Octal | 0.1230 | 8.13 (+/-0.03) | 2015.00 | 17.3 |
| | 0.1182 | 8.46 (+/-0.03) | 2015.81 | 17.0 |
| | 0.1133 | 8.83 (+/-0.04) | 2016.97 | 10.1 |
| | 0.1064 | 9.39 (+/-0.04) | 2013.85 | 5.9 |
| Decadal | 0.0996 | 10.04 (+/-0.05) | 2010.33 | 30 |
| | 0.0948 | 10.55 (+/-0.06) | 2012.07 | 52 |
| | 0.0908 | 11.01 (+/-0.07) | 2013.98 | 57 |
| | 0.0840 | 11.90 (+/-0.1) | 2014.92 | 30 |
| Decadal-Plus | 0.0763 | 13.10 (+/-0.1) | 2013.77 | 12.5 |
| | 0.0718 | 13.92 (+/-0.1) | 2014.34 | 10.4 |
| | 0.0664 | 15.05 (+/-0.1) | 2014.82 | 7.5 |

**3.2 Forecasting the future variation in SSN.** A forecast was obtained by fitting sinusoids to each of the eleven components obtained by INF and shown in Figures 2 A, B and C. The sinusoids were fitted at solar cycle 24, around year 2012, as follows. The amplitude was measured as the peak height of each component. The phase was estimated as the average of the zero crossing times at solar cycle 24 for each component. For example, the sinusoid fit to the 10.04 year period component was $30\cos(2\pi(t - 2010.33)/10.04)$. The period, frequency, phase and amplitude of the fitted sinusoids are listed in Table 1. As amplitudes of the decadal group exceed by several times the amplitudes of the octal and decadal-plus groups, destructive interference between the projected decadal group sinusoids determines when grand minima occur. The projections of the decadal group of sinusoids between year 0 and 2500 are shown in Figure 3A along with the sum of the projections. The forward projection, from 2012 to 2500 shows a grand minimum labelled as "Next" in Figure 3A. Also evident in Figure 3A are a sequence of grand minima in the time interval extending back to year zero AD. The back-projected grand minima vary in duration from about 100 years to about 200 years. The grand minima are separated by grand maxima of duration in the range 50 to 100 years. This result is consistent with the statistical analyses of grand minima/maxima duration in reconstructed SSN, (Usoskin et al 2007, Inceoglu et al 2015). The grand minima evident in the back projection, labelled Oort, Wolf, Sporer, Maunder and Dalton in Figure 3A, coincide reasonably closely with centre dates of grand minima as assessed from reconstructed sunspot number and listed in Usoskin et al (2007) and Inceoglu et al (2015). The centre dates of listed grand minima are indicated by the vertical reference lines in Figure 3A. While the grand minima correspondence may just be random coincidence, here we take it as evidence that the reliability of the method is acceptable over the centennial range for grand maxima/minima occurrence.

The projection procedure above was applied to the components in the octal and decadal-plus groups and similar sinusoidal projections to that in Figure 3A were obtained. The individual octal and decadal-plus component projections are not shown here separately. However, the sinusoidal fits obtained are listed in Table 1. The summed projections for each group are shown in Figure 3B between 1750 and 2200. All three group projections show an extended grand minimum commencing around 2030 and lasting until about 2150. Evidently destructive interference within the octal group and within the decadal-plus group occur at about the same times as destructive interference in the decadal group, i.e. the octal, decadal and decadal-plus variations exhibit a similar long term amplitude variation. This is a characteristic of the time variation in amplitude of sideband components produced by a single amplitude modulation of a carrier, suggesting that the octal and decadal-plus periodicities may have a common origin, for example, as higher and lower sidebands resulting from amplitude modulation of one or more of the decadal group of periodicities. While the decadal projection dominates the octal and decadal-plus projections during grand maxima, for example, during the Modern Maximum between 1950 and 2000, in the grand minima, for example the Dalton minimum, ~1810, and in the projected Next grand minimum, ~2030 to ~2150, the octal, decadal and decadal-plus projections are of similar amplitude. The result is that, during grand minima, octal or decadal-plus periodicity occasionally dominates decadal periodicity, consistent with recent observations during the Maunder grand minimum, Yan et al. (2023). Similarly, during the last seven grand minima Inceoglu (2024) found that solar cycle lengths during these grand minima exhibit a bimodal distribution with local maxima clustered around 7 and 13 years. Similarly, from observations during grand minima in the first millennium BCE, for example, during the 300 – 400 BCE grand minimum, decadal periodicity is dominated by octal and decadal-plus periodicity, Brehm et al (2025). Consistent with the results of Figure 3, Saha and Nandy (2024), based on dynamo simulation studies, found that "*the 11-yr sunspot cycle disrupts to some extent, and power distribution in different periodicities considerably re-organizes during grand minima. Several shorter and longer periodic phenomena become prominent, as elucidated by Saha et al. (2022)*."

Figure 3C shows the sum of all three group projections between 1600 and 2200. The time of peak solar cycle 24, the time at which the eleven sinusoids were fitted, is indicated by the dotted reference line. The back projection from this date reproduces the Maunder and Dalton grand minima reasonably well in terms of occurrence and duration, (see Figure 1, Usoskin (2023)). The forward projection forecasts a grand minimum in SSN from about 2030 to 2150 labelled "Next".

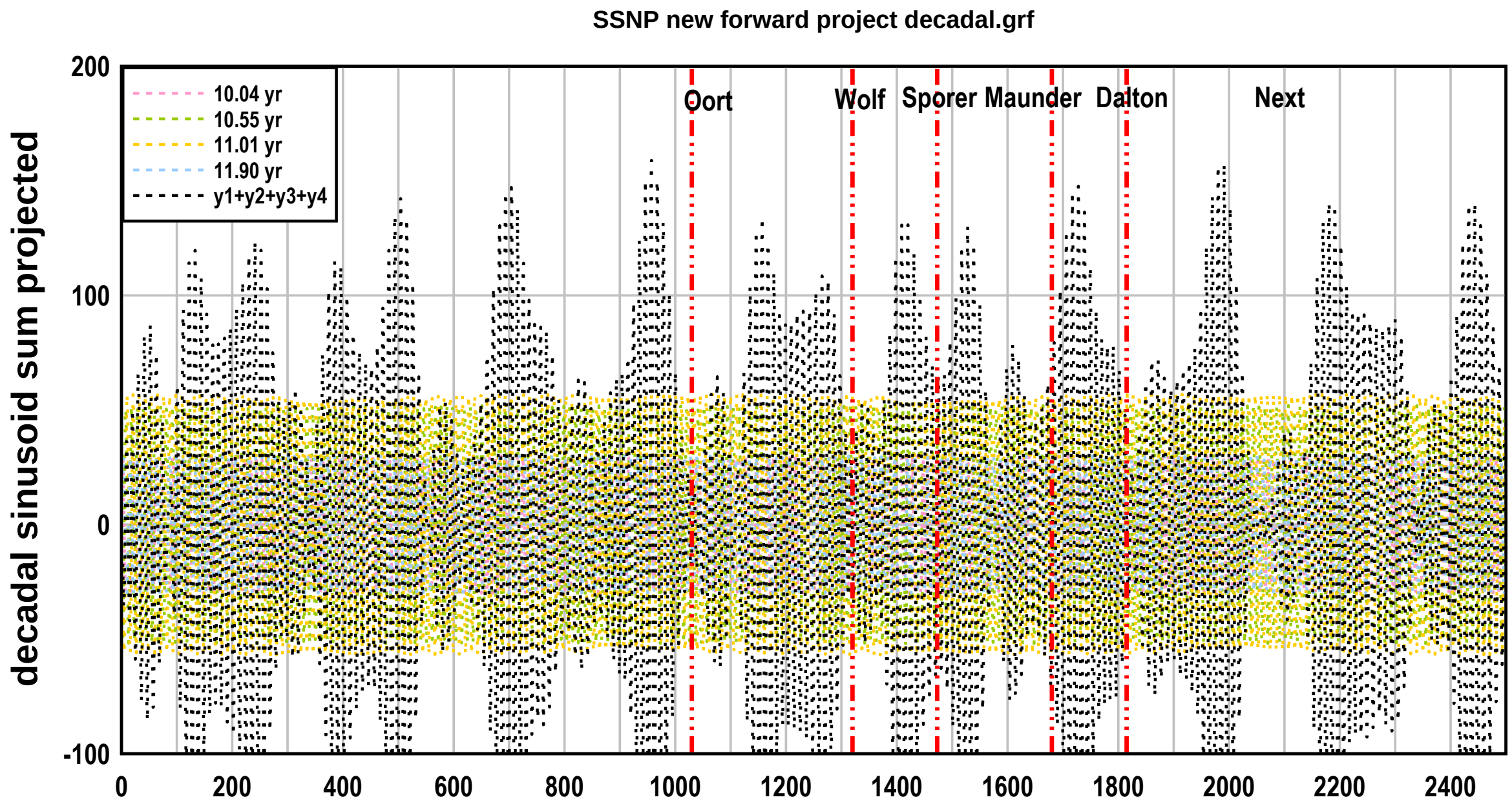

(A)

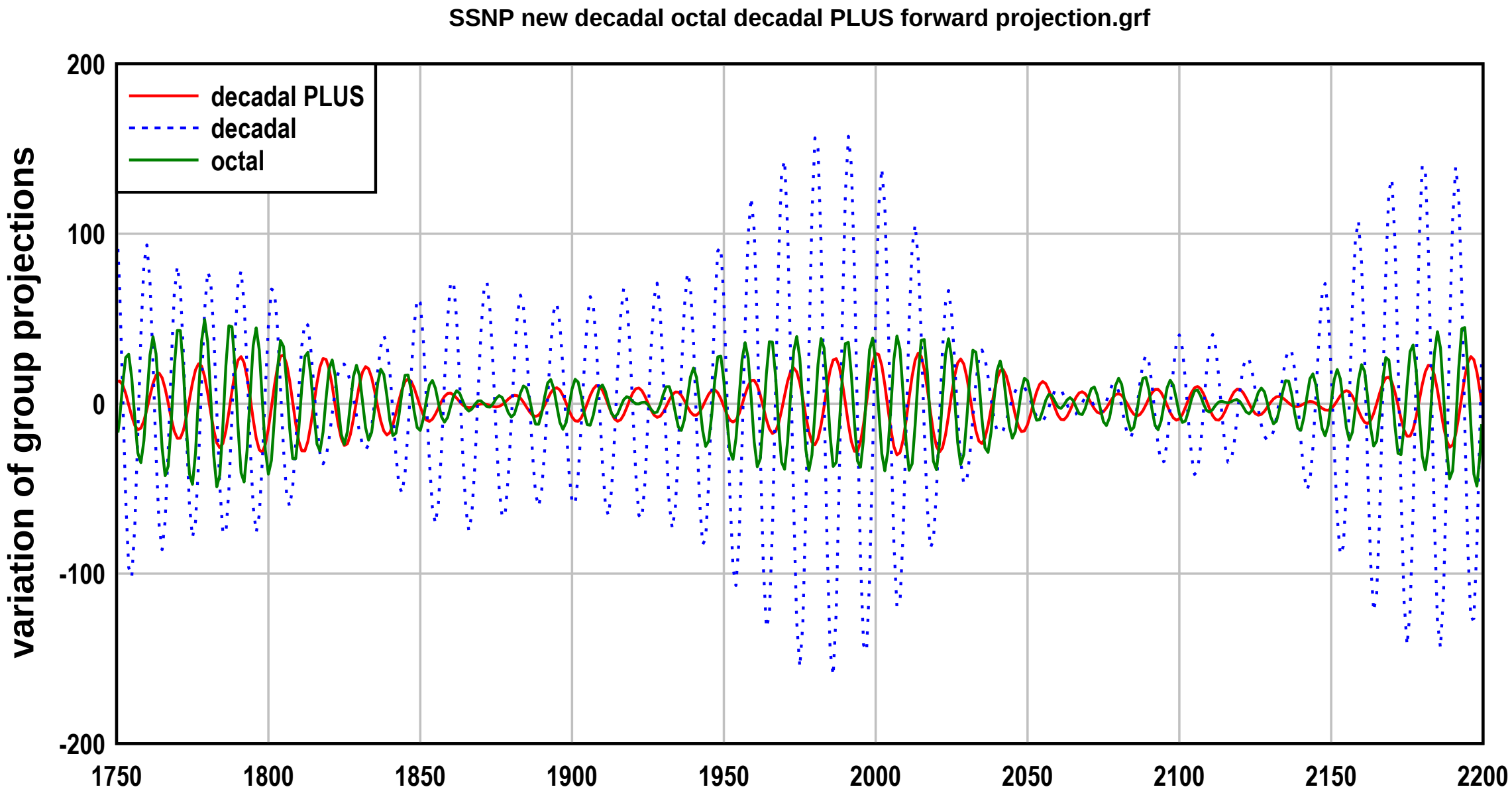

(B)

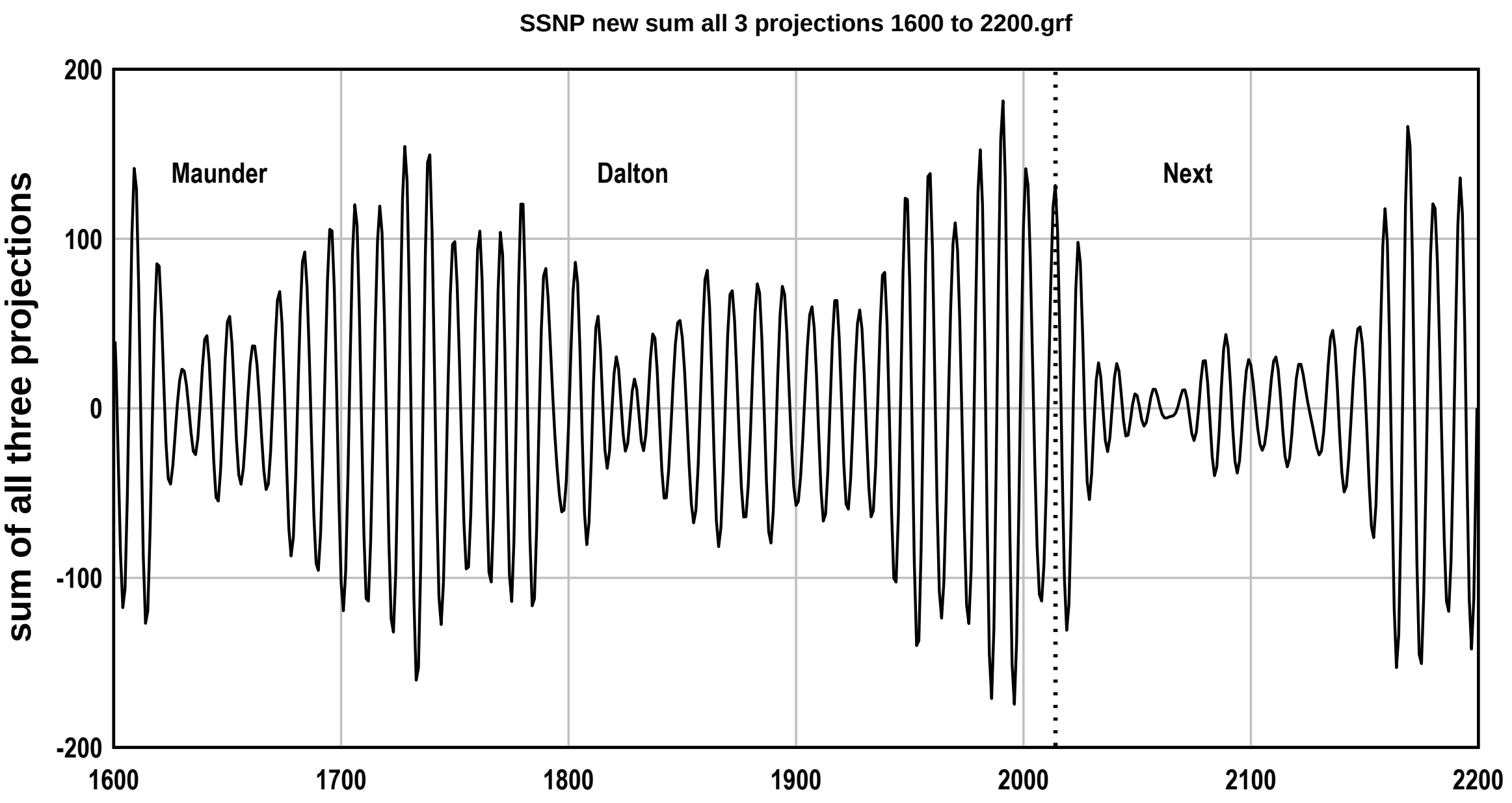

(C)

**Figure 3. (A) The projection of sinusoids fitted to the decadal group components at solar cycle 24 (~ 2012) back to year 0 and forward to year 2500 generates, by destructive interference, a sequence of grand minima. In the backward projection, 2012 to 1000, the generated grand minima coincide fairly closely with grand minima in reconstructed SSN records, with centre dates of reconstructed grand minima, Usoskin et al (2021), indicated by labelled reference lines. In the forward projection, 2012 to 2500, a grand minimum labelled "Next" occurs between 2030 and 2150. (B) Shows the projection of each group sum from 1750 to 2200. During grand minima the octal and/or decadal-plus contributions occasionally exceed the decadal contribution, e.g. during 2040 to 2070, also during the Dalton Minimum, 1810 to 1830. (C) The projection of all three group sums 1600 to 2200 reproduces, in the back projection, the Maunder Minimum, between 1600 and 1700, and the Dalton Minima, ~1820, and in the forward projection forecasts the Next grand minimum, 2030 to 2150.**

The sinusoids fitted to components at 2012 when projected back to 1600 and summed reproduce quite closely the occurrence time and duration of the Dalton Minimum and the Maunder Grand Minimum, Figure 3C. This is evidence of the persistence of the period and phase of the components obtained by the INF method over intervals of a few hundred years. It is, therefore, reasonable to expect that sinusoids fitted to components at 2012 and projected forward from 2012 to 2200 would reliably forecast the occurrence of grand minima in this time interval. As Figure 3 shows the forward projection forecasts a grand minimum from about 2030 to 2150 labelled "Next" in Figure 3. Evidently, as Figure 3B shows, the grand minimum is primarily due to destructive interference of the four decadal projected sinusoids. Due to the complex interference between the three projected group sinusoid sums during a grand minimum (see Figure 3B) it is unlikely that the solar cycle-to-solar cycle variation during the projected grand minimum is accurate. However, some features within the forecast grand minimum may be of interest for comparison with other studies e.g. Velasco Herrera et al (2011, 2015), who also forecast a grand minimum during the current century. Accordingly, the detail of the projected grand minimum is reproduced in Figure 4.

Important features of the predicted grand minimum are as follows. Solar cycles 26, 27, 28, 29 and 30 are forecast to be very small relative to the amplitudes of solar cycles 21, 22 and 23 observed during the Modern grand maximum. Also the solar cycle lengths in solar cycles 26 – 29 may be quite short, about 8 years on average, due to the dominance in this interval of the octal group of components. Towards the end of the forecast grand minimum, 2120 to 2150, the lengths of solar cycles 34 to 37 could be quite long, about 12 years, due possibly to the reassertion of dominance by the decadal and decadal-plus components relative to the octal component in this interval. Figure 4 is a forecast of the relative strength of the solar cycles. Forecasting the SSN involves establishing the amplitude ratio between observed SSN, Figure 1A, and the forecast solar cycles in Figure 4 and using the ratio to find the SSN of forecast solar cycles. For example, Figure 4 forecasts a trough-peak height of 80 for solar cycle 26. Compared with the back projection trough-peak height for cycle 23 of 315 in Figure 4, the amplitude of cycle 26 is just one quarter the amplitude of cycle 23. In the observed SILSO SSN record, Figure 1A, the trough-peak amplitude of cycle 23 was 162 so the forecast maximum SSN of solar cycle 26 is 162/4 = 40. We note that Rodriguez et al (2024 a,b) predicted a solar cycle 26 maximum about 120. Cao et al (2024), from one of their models, forecast a maximum of about 120 and another model, the Transformer model, forecast a maximum of about 50. Scafetta (2012) developed a model of SSN with three decadal cycles, (periods: 9.93, 10.87 and 11.86 years), to forecast a Dalton-like minimum between 2000 and 2050. De Jager and Duhau (2012) stated categorically that “*No Grand Minimum is expected to occur during the millennium that has just started*”. Solanki et al (2013) stated that “it is equally likely that the next grand extremum will be a grand maximum as a grand minimum”. A list of seven recent predictions of peak SSN for solar cycle 26 is provided by Abdel Rahman and Eid (2026). The minimum is 118 and the average is 160, four times the forecast maximum, 40, of this paper. Asikainen and Mantere (2023) predicted a cycle 26 stronger than cycle 25 and an increasing trend in solar activity for the next decades. The forecast, in this paper, of a precipitous decline to a peak of 40 for cycle 26 is supported in Section 4.2 by a “shah” model of grand minima formation and by wavelet analysis of reconstructed SSN, Figure 8, which shows abrupt onsets of wide grand minima. The effect of solar irradiance variability on climate is reviewed by Solanki et al (2013) and the effect of a grand solar minimum on climate would be small according to modelling, (Shindell et al 2001, Feuler and Rahmsdorf 2010, Jones et al 2012, Rozanov et al 2011, Anet et al 2013, Ineson et al 2015, Maycock et al 2015). See also de Jager and Usoskin (2006).

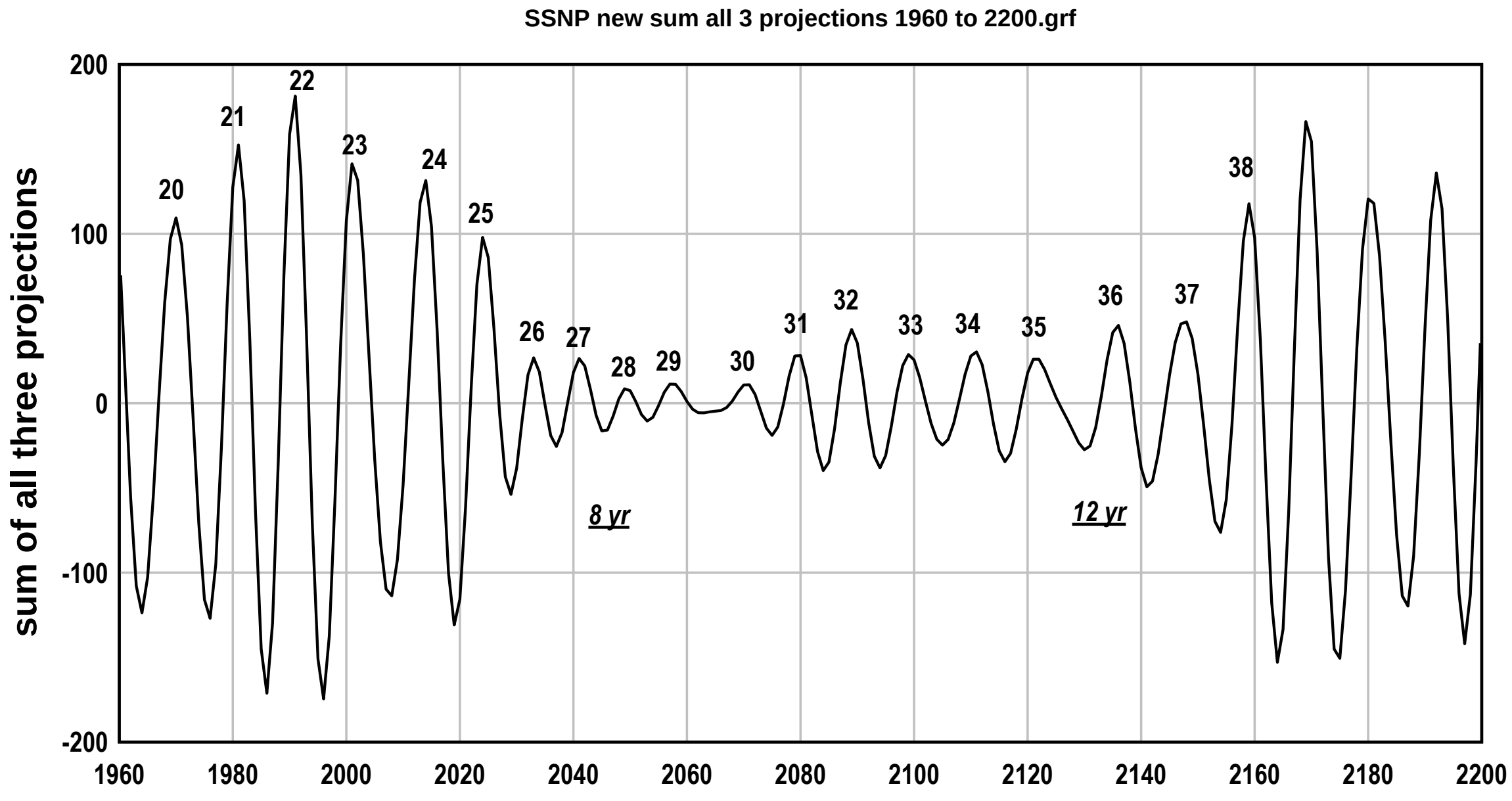

**Figure 4. (A) High resolution forecast from solar cycle 24 back to 1960 and forward to 2200. The forecast cycle SSN amplitudes can be estimated by comparing the trough - peak SSN of solar cycle 24, (known SSN amplitude 174) with the trough – peak number of a forecast solar cycle. Note that the spectral content change from ~8 yr periodicity to ~12 year periodicity during the grand minimum is consistent with simulations, Saha et al (2022), and observations, Inceoglu (2024), of solar dynamo behaviour.**

**3.3 The Waldmeier Effect.** This paper has separated the time variations of the octal, decadal and decadal-plus periodicities in SSN, see Figure 2. The Waldmeier Effect is an anti-correlation between solar cycle amplitude and length of the ascending phase of the solar cycle (Waldmeier 1935, Takalo and Mursula 2018). As evident in Figure 4, the combination of the three group projections during a grand minimum can result in extreme variations in solar cycle length and amplitude as dominance between the three groups changes.

Solar cycle-to-solar cycle variations are also present during a grand maximum but, due to the dominance of the decadal group, are less extreme. As it happens, the records for the studies of the Waldmeier Effect, for example, 1755 to 2009, Takalo and Mursula (2018), are in time intervals dominated by the decadal group of components, see Figure 3B. The only exception, which is brief, is around 1810 during the Dalton Minimum. The objective here is to show that the Waldmeier Effects, reported in earlier studies, can be accounted for by interference between the octal component sum, Figure 2B, and the decadal component sum, Figure 2A.

Figure 5 shows the variation of the sums during the interval 1900 to 2024. Some of the effects immediately noticeable are as follows: At the start of solar cycle 20 in Figure 5 the decadal component is rising at the same time the octal component is falling so the combined effect is a long rise time or low rate of rise of the SSN. At the peak of cycle 20 the decadal and octal components are in almost exact anti-phase tending to reduce the combined SSN peak amplitude. Thus in cycle 20, an inverse relationship between rise time and cycle amplitude occurs (i.e. when the rise time is high the cycle amplitude is low).

Cycle 22 presents the converse situation: The decadal component is rising at the same time the octal component is rising so the combined effect is a short rise time or a high rate of rise of SSN. At the peak of cycle 22 the decadal and octal components are in phase and tending to increase the combined peak amplitude of the SSN. Thus also in cycle 22 an inverse relationship between rise time and cycle amplitude occurs (i.e. when the rise time is low the cycle amplitude is high). Thus, the interference of the decadal and octal components provides an explanation of the Waldmeier Effect, the observation that the rise time of a sunspot cycle varies inversely with the cycle amplitude, i.e., strong cycles rise to their maximum faster than weak cycles, (Svalgaard and Hathaway 2020, Cameron and Schussler 2008a, Takalo and Mursula 2018, Petrovay 2020).

Whenever the octal component is in-phase/out-of-phase to the decadal component, the length between solar cycles decreases/increases. For an example in Figure 5, the in-phase condition at solar cycles 18 and 19 leads to a short cycle length, 10 years, between the two cycles. The out-of-phase condition between solar cycles 23 and 24 leads to a long solar cycle length, 12 years, between the two solar cycles. The result is an inverse relationship between cycle amplitude and the length of the preceding cycle - another Waldmeier Effect, Hathaway et al (1994). Unusual solar cycles are discussed by Nandy et al (2011) and Ahluwalia and Jackiewicz (2012). A simple model of the Waldmeier Effects based on interference between octal and decadal components is presented in the Discussion.

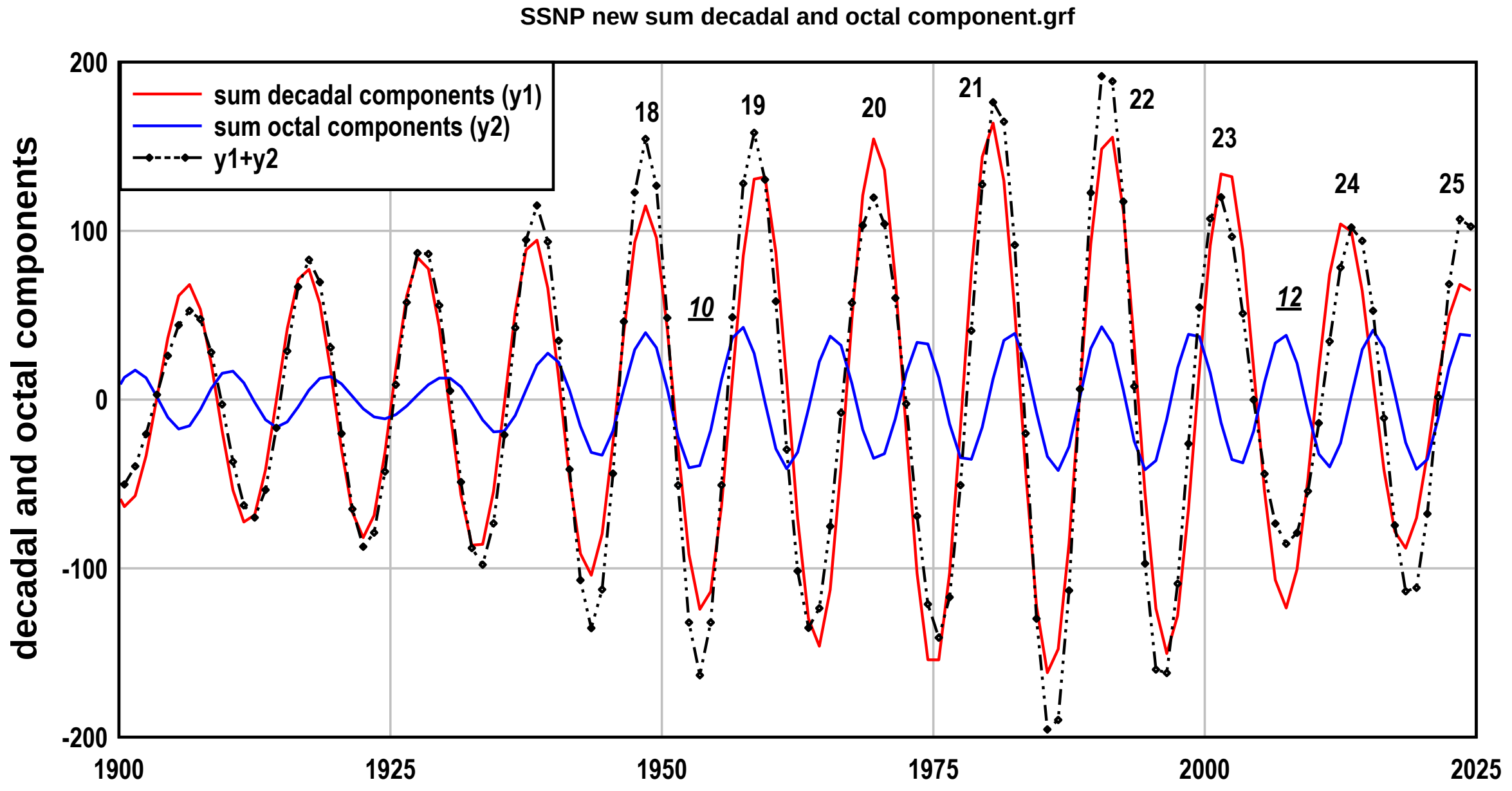

**Figure 5. Solar cycles 20 and 22 are illustrative of how the octal component sum interferes with the decadal component sum to generate the Waldmeier Effect, as discussed in the text.**

**3.4 A forecast grand minimum based on long period cycles in reconstructed SSN.** The objective in this analysis was to apply the method of component estimation by INF and sinusoid fitting to low frequency components of reconstructed SSN to obtain a low resolution SSN forecast. Komitov and Kaftan (2003), applied a similar method to the Schove series, (Schove, 1955), to obtain a low resolution SSN forecast. The Schove series provides the variation in SSN from 296 to 1955 at solar cycle (~ 11 year) resolution. Periodogram analysis identified the main long term periodicities as 100, 122, 205, 353 and 1220 years. By fitting and projecting five sinusoids at these periods, Komitov and Kaftan (2003) forecast a grand minimum, falling from the peak of the Modern grand solar maximum at ~1960 to a lowest amplitude solar cycle at ~2070, with a following grand maximum at ~2160. Their forecast is broadly similar to the duration of the high resolution forecast grand minimum obtained above by projection of the decadal-range periodicities derived from SILSO SSN (see Figure 4A).

Reconstructed annual SSN is available for the interval 971 to 1899, Usoskin et al (2021). To define more closely the long-term periodicities the SSN anomaly from this record was padded with zeroes to 2048 years and the periodogram found (Figure 6A, diamond symbols). Four major long-term components were identified, at 60, 125, 212 and 768 years as indicated by the vertical reference lines. The four components at these periods were obtained by the INF method at 5% bandwidth. The four components were summed and, with the sum padded to 2048 years, a periodogram was obtained (Figure 6A, blue squares). This indicated an underestimation of the component amplitude by the INF method by the factors labelled in Figure 6A, i.e. 1.09, 1.34, 1.53 and 2.90. Sinusoids were fitted to the INF components at around year 1830, the amplitudes of the fitted sinusoids were increased by the previously determined factors, the sum of the sinusoids was obtained and the periodogram of the sum found, Figure 6A, (red * symbols) confirming an adequate amplitude adjustment.

As expected the sum of the four projected sinusoids replicates closely the component sum in the interval 971 to 1899 (Figure 6B, label y2+y3+y4+y5). The forward projection from 1899 forecasts a minimum between 1910 and 1940, (the Gleissberg Minimum), and a grand maximum, (the Modern Grand Maximum), extending from about 1960 to about 2030. This provides a validation of forecasting by the method for at least 120 years, the time interval from 1899 to the present. The forecast Modern maximum is followed by a broad minimum extending from about 2030 to about 2200, labelled “Next” in Figure 6C. This forecast minimum is broadly consistent with the result in section 3.2 where projection of the decadal-range components of SILSO SSN forecast a wide grand minimum in SSN between 2030 and 2160, and, similarly, is broadly consistent with the forecast of Komitov and Kaftan (2003) of a minimum in SSN between 1960 and 2160.

It is interesting that the back projection to the interval 0 to 971, Figure 6C, generates two grand minima that coincide quite closely with the two wide grand minima identified in reconstructed SSN in this time interval, (Inceoglu et al 2015, Usoskin et al 2025, Temaj et al 2026), at 676 and 417 years, and marked, in Figure 6C, by the pink reference lines.

There are seven periodicities indicated in the periodogram shown in Figure 6A, between 40 and 100 years. Selecting just the strongest, the 60 year periodicity, for analysis with the longer (> 100 year) periodicities results in a projection that emphasises wide grand minima such as the Sporer or the Gleissberg and diminishes the contribution to the projection of narrow minima such as the Dalton. However, the objective here was to show that a forecast based on low frequency periodicity generates a forecast similar to that based on the projection of higher frequency, decadal-range, periodicity (i.e. a forecast of a broad minimum following the Modern Maximum in SSN).

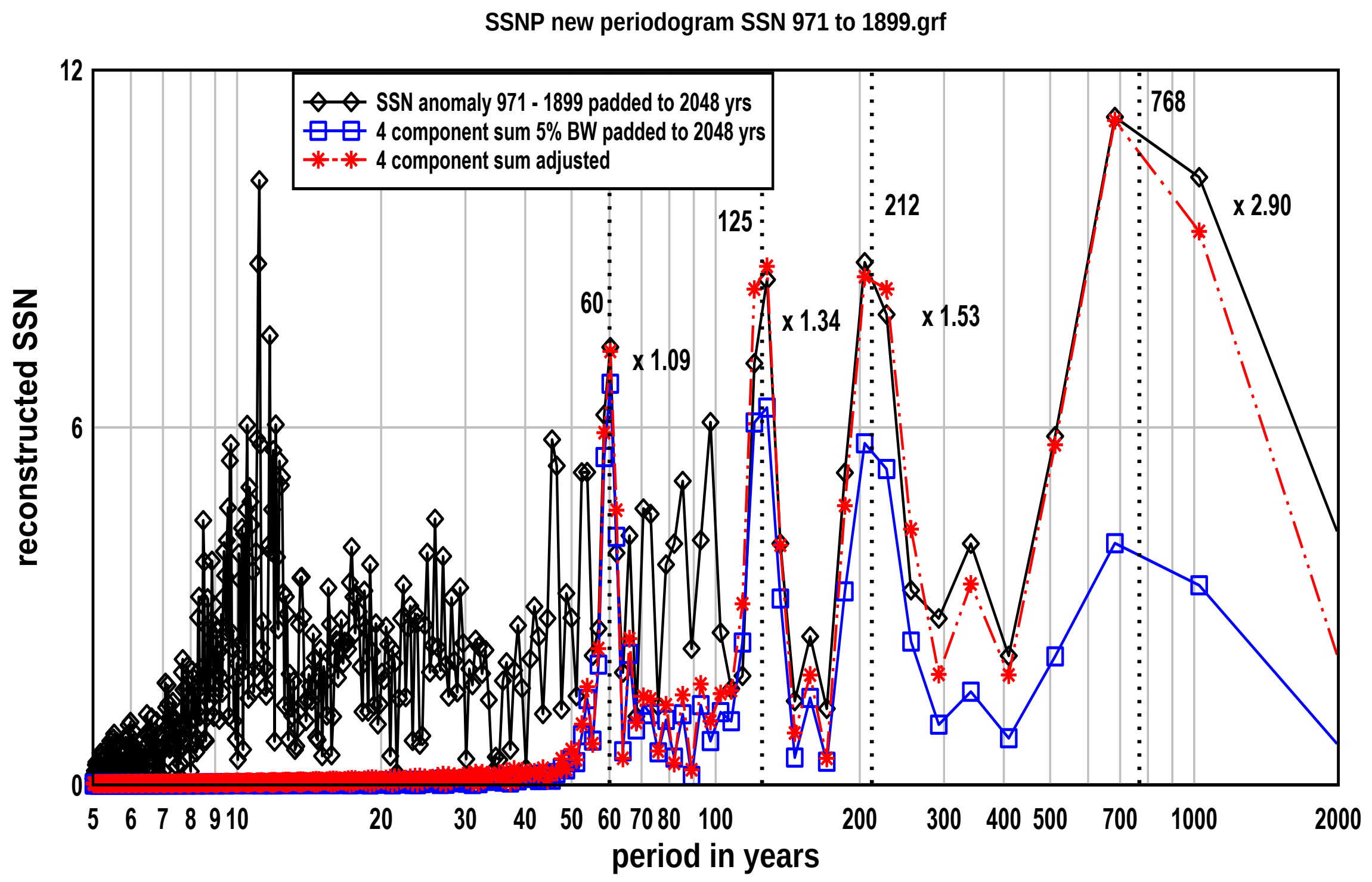

(A)

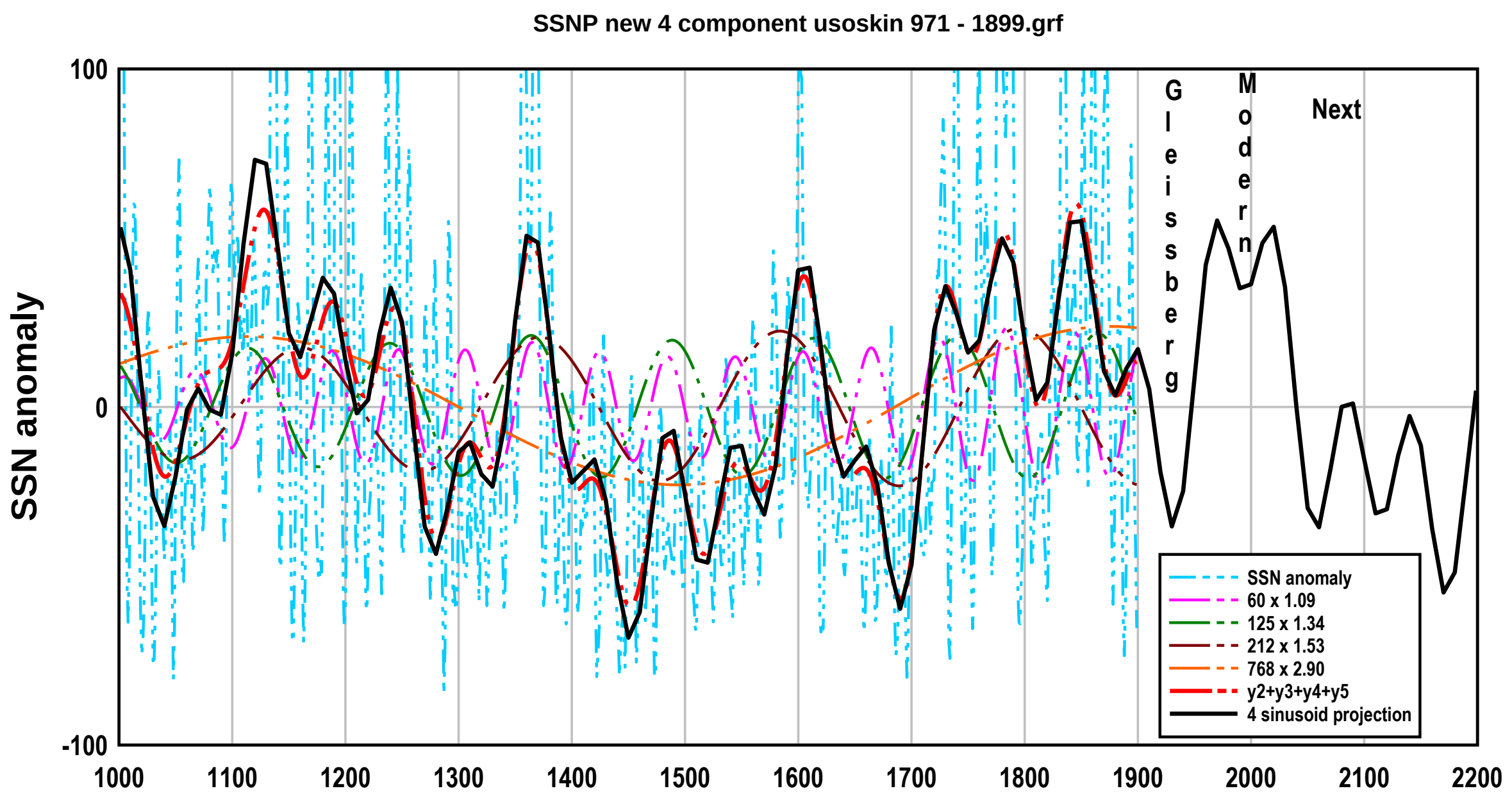

(B)

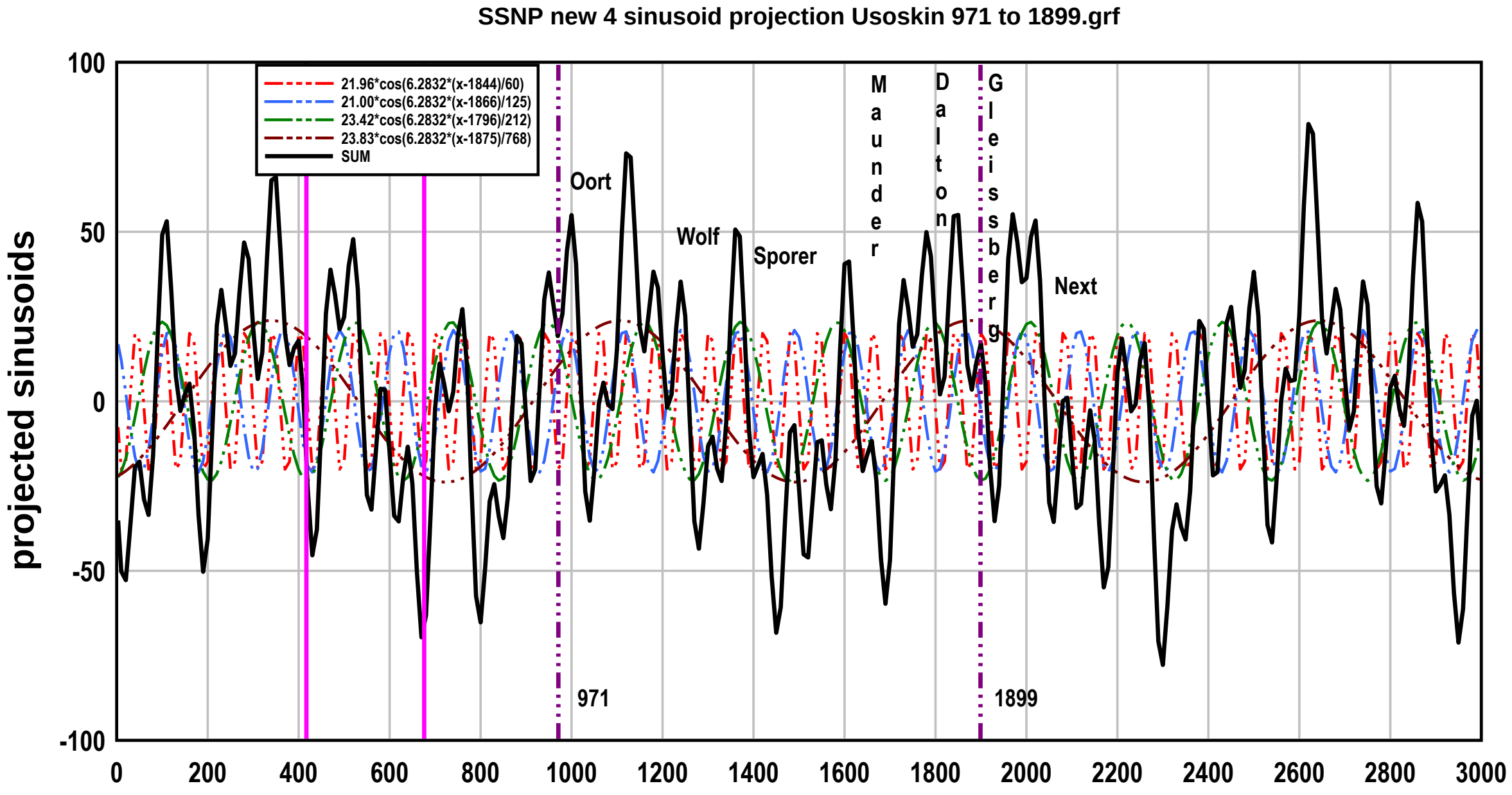

(C)

**Figure 6. (A) A periodogram obtained by FFT of the anomaly of the SSN reconstruction 971 – 1899, Usoskin et al (2021), (diamond symbols). The four strongest long-term periodicities are identified with vertical reference lines at 60, 125, 212 and 768 years, the latter period obtained by curve fitting software. The periodogram (square symbols), is obtained from the sum of INF components at the four periods. The INF components underestimate the amplitudes at the four periods by the factors, 1.09, 1.34, 1.53 and 2.90 as labelled. When the component amplitudes are adjusted by these factors the periodogram, (asterisk symbol), is obtained. (B) Shows the SSN anomaly, the amplitude-adjusted components, the sum of the components and the sum of the projected sinusoids fitted to the adjusted components. The four-sinusoid projection, from 1899, forecasts a minimum corresponding to the Gleissberg minimum, the Modern grand maximum and a following broad minimum. (C)**

**Shows the four-sinusoid projection 0 to 3000 forecasting the Next grand minimum extending from about 2030 to 2200. The pink reference lines indicate the occurrence times of the two wide grand minima in the first millennium as obtained from reconstructed SSN, (Inceoglu et al 2015, Usoskin et al 2025, Temaj et al 2026).**

## 4. Discussion.

It is a peculiarity of review articles on solar activity and on the solar dynamo, (Ossendrijver 2003, Balogh et al 2014, Clette et al 2014, Hathaway 2015, Charbonneau 2014, Charbonneau 2020, Biswas et al 2023, Saha and Nandy 2024), that the reviews usually open with illustrations of the time variation in SSN, similar to Figure 1A, but never illustrate, to the best of our knowledge, the variation of SSN in the frequency domain, similar to Figure 1B. This is despite the fact that the time and frequency domains contain and provide different but complementary insights into the same phenomenon. After perusal of the reviews on solar activity referenced above, a researcher would be unaware that the variation in solar activity is characterized by a few well defined frequency components. An exception is the recent article by Dash et al (2023) where an effort is made to compare observational solar activity spectra with modeled spectra during intervals of grand minima and maxima. The major observation of the current work is that the long-term variation in solar activity is largely described by interference between the components identified by spectral analysis as shown in Figure1B and Figure 6A. The time variation of the ~decadal-range components (Figures 2A, 2B and 2C), suggests that the phase of a component is persistent whereas the amplitude of a component slowly varies on a centennial scale. Persistent cycles in solar activity have been discussed by Gil-Alana (2009) and Maddanu and Proietti (2022) with their analysis suggesting the presence of “stationary cyclical long memory” in the sunspot-generating process.

**4.1 Forecasting the next grand minimum.** The basic assumption adopted in the projection of sinusoids to obtain a forecast is that the amplitude and period of the components that the sinusoids represent remain essentially constant over the forecast interval. A slow variation in amplitude is evident in the components shown in Figure 2A. For example, the amplitude of the decadal components roughly doubles between 1700 and 2024. This slow variation implies that, for the forecast based on ~decadal-range periodicity, the sinusoidal projection of amplitude should be reliable for a range of about five solar cycles (i.e. about 60 years) from the time used for fitting the sinusoids. Another important factor is the accuracy of periods obtained by Fourier analysis (Figure 1B). A 1% error in a ~decadal period, for example 10.1 years instead of 10.0 years, implies that the two corresponding projections would be in anti-phase after 500 years. This suggests that the phase interference that leads to grand minima should be reliable to about 100 years from the time point at which sinusoids are fitted. This is supported by the accuracy of back projections to prior observed grand minima. For example the back projection of the ~decadal-range sinusoids from the fitting point at solar cycle 24, ~2012, reproduces quite closely the solar cycle variation at solar cycles 5 and 6, in the Dalton Minimum around 1815, (Figure 3C). Beyond about fifty years this type of projection will become unreliable in respect to individual solar cycle parameters but may be useful for about one hundred years to indicate long-term variation in solar activity.

The low-resolution forecast, Figures 6B and 6C, based on centennial-scale periodicity in reconstructed SSN record that ended in 1899, accurately forecast the Gleissberg Minimum around 1920 and the Modern Grand Maximum, about 100 years ahead, and may be accurate up to 500 years ahead (c.f. Steinhilber and Beer 2013).

Explanations of the long-term variation in solar activity based on dynamo models are reviewed by Karak (2024) and by Charbonneau (2020) who observed: *"Since the basic physical mechanism(s) underlying the operation of the solar cycle are not yet agreed upon, attempting to understand the origin of the observed fluctuations of the solar cycle may appear to be a futile undertaking"*. It is interesting therefore, to see how a model with a few spectral components that are closely and evenly spaced in frequency, as indicated in Figure 1B, can produce the type of complex, short- and long-term variations that are characteristic of the long-term variation in the SSN record.

**4.2 A model of grand minima formation.** The *shah* symbol, III(x), represents a function comprised of a series of delta functions spaced at regular frequency intervals or time intervals in the respective domains. The Fourier transform of a *shah* function of spacing $\Delta$f in the frequency domain is a *shah* function of spacing $\Delta$t = 1/$\Delta$f in the time domain. The decadal group of peaks shown in the periodogram of Figure 1B approximates a four component *shah* function of average spacing $\Delta$f = 0.0059 year$^{-1}$ in the frequency domain. The octal and decadal-plus groups occur at similar but smaller spacing, 0.0049 and 0.0045 year$^{-1}$ respectively, but with peaks of much lower amplitude. If the amplitudes of the spectral components within each group are nearly equal, the Fourier transform of each group into the time domain will approximate a shah function in the time domain with the following spacing: decadal group: 1/0.0059 = 170 years, octal group: 1/0.0049 = 204 years, decadal-plus group: 1/0.0045 = 222 years. On the basis of this we would expect the long-term variation in SSN to be characterized by an approximately 200 year periodicity between grand minima.

To model the expected variation of SSN in the time domain we generated four co-sinusoids of approximately decadal period and of equal frequency spacing 0.005 years$^{-1}$, Figure 7A. The objective is to illustrate that as the components move in and out of phase or if the relative amplitudes of the components vary slowly so that sometimes all the amplitudes are nearly equal and sometimes just two components dominate, the result in the time domain is a variation that can include a wide variety of different types of grand maxima and grand minima. The model, nominally the "shah" model, in Figure 7A has a mean frequency of 0.0925 years$^{-1}$. This results in an average short-term variation of period ~ 10.8 years in the time domain with extended grand minima separated by 200 years, Figure 7B. This type of variation would represent a prolonged grand minimum such as the Sporer minimum. Figure 7B also shows how different amplitude configurations of the four spectral components produce different types of grand minima as the relative amplitude of the contributing components change to provide, in the frequency domain, the following amplitude configurations: IIII, .II., I..I . Comparing the result in the time domain of different component configurations in the frequency domain, Figure 7B, shows that the appearance of Dalton-like grand minima in the time domain, (spacing $1/\Delta f$ = 200 years), can originate from amplitude configurations like .II., or configurations like I..I, (time domain spacing $1/3\Delta f$ = 67 years), or configurations like .I.I, (time domain spacing $1/2\Delta f$ = 100 years; not shown in Figure 7B).

The long-term periodicities associated with the model emerge in spectral analysis only if the time variations in Figure 7 are rectified, i.e. only if the positive half of the model variation is taken as being representative of SSN. The model illustrated in Figure 7 suggests that, provided the amplitudes of the components vary slowly, as the components move in and out of phase interference between the four decadal components, (Figures 1B and 2A), can generate a wide variety of types of grand minima and maxima that occur at intervals consistent with the duration of grand minima and maxima as assessed by waiting time distribution analysis of reconstructed SSN records, (Usoskin et al 2007, Inceoglu et al 2015). The interference between octal, decadal and decadal-plus components would increase the possible range and complexity of long-term variability. An example of complexity due to interference between components of non-equal amplitude is Figure 3A where no long-term repetitive pattern of grand minima/maxima is evident. The deterministic process described here contrasts with the view that grand minima episodes are caused by some type of stochastic process in the Sun that "triggers" the normal, single frequency, mode of a solar dynamo process into an abnormal and weaker mode, Charbonneau (2020). This type of process would generate a frequency spectrum of SSN in the decadal range comprising a single strong peak at the dynamo frequency and a broad band of weak sidebands. Constructive and destructive interference between two dynamos with slightly different frequencies has been studied as a means to generate grand solar maxima and minima (Shepherd et al 2014, Ölçek et al 2019). This dual dynamo process would generate a spectrum comprising strong peaks at the two dynamo frequencies and a time variation similar to the variation labeled Dalton-like in Figure 7B. Both types of spectra are contrary to the observed spectrum (Figure 1B), which clearly contains four strong decadal peaks. To produce wide grand minima with narrow grand maxima, as in Figure 7A, three or more approximately equal amplitude components are required.

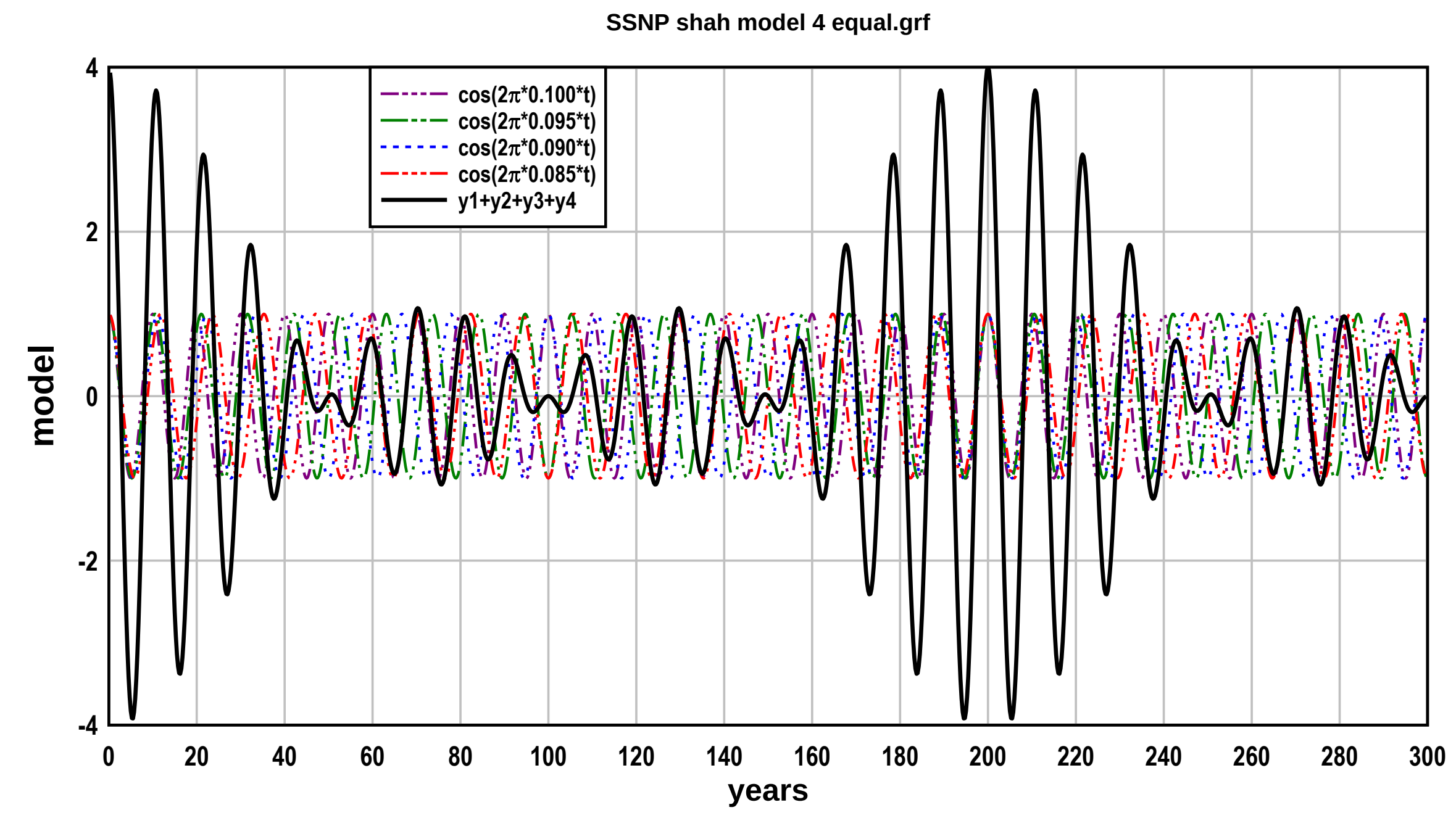

(A)

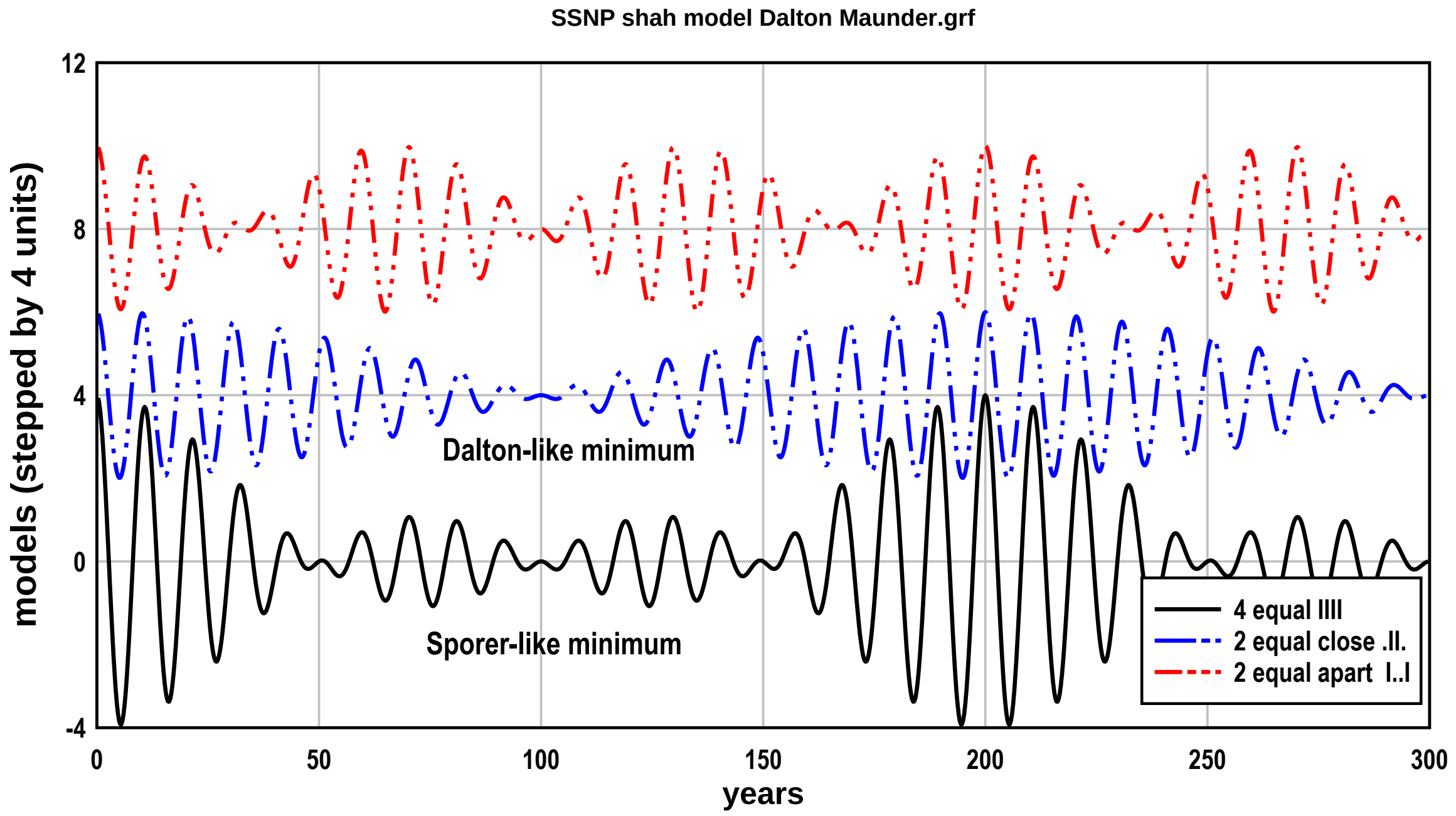

(B)

**Figure 7. (A) Shows the variations of four equal-amplitude sinusoids of different frequency, broken lines. The frequencies are equally spaced, $\Delta f = 0.005$ year$^{-1}$, in the frequency domain, see equations in the label. The sum, full line, leads to grand maxima, equally spaced by $\Delta t = 200$ years, in the time domain as discussed in the text. (B) Different types of grand minima, Maunder-like or Dalton-like, arise from different amplitude configurations of the four sinusoids in the frequency domain. This model suggests that, if the amplitudes of the four interfering sinusoids varied slowly with time, a wide range of different types of grand minima and maxima would develop over a long time interval.**

An important feature of the “shah” model is that when the amplitude of succeeding solar cycles is trending sharply downwards, for example during years 20 to 40 or 220 to 240 in Figure 7A, this sharp downward trend in cycle amplitude leads into wide Sporer-like grand minima. Conversely, if the downward trend of solar cycle amplitude is more gradual, as illustrated by the middle curve in Figure 7B, a narrow Dalton-like grand minimum results. Consistent with this is the work of Lockwood et al (2011) who found that a steep decline in solar cycle amplitude was a precursor for the prolonged Maunder Minimum, whereas a gradual decline in solar cycle amplitude led to the shorter Dalton Minimum.

In recent times the trend in solar cycle amplitude has been sharply downwards (see Figure 1A here and Lockwood et al (2011)), suggesting, from the shah model here and the analysis of Lockwood et al (2011) that a wide Sporer-like grand minimum will follow. A Sporer-like or Maunder-like grand minimum was predicted by the projection analysis detailed in Section 3 where destructive interference between projected decadal range components forecast a grand minimum 2030 to 2150, Figure 4. Abreu et al (2008) predicted that the Modern Grand Maximum would end within 30 years but were unable to specify the type of ensuing grand minimum. One view is that the occurrence of grand minima is not the result of interference between long term cyclic variations, but is instead a special state of the solar dynamo and that, once triggered into a grand minimum by some stochastic/chaotic process, the dynamo is “trapped” in this state and its behavior is then driven by “deterministic intrinsic features”, (Usoskin 2023, Karak 2024). As mentioned earlier de Jager and Duhau (2012), based on a projection of the ~2400 year period Hallstaatt cycle in cosmic ray flux, Usoskin et al (2016), forecast no Grand Minima occurring in the current millennium, 2000 – 3000. However, there is evidence that the Hallstaatt cycle may be due to geomagnetic rather than solar magnetic variability, Nilsson et al (2024).

Charbonneau (2020) observed “*At this writing we still do not know what triggers Grand Minima, or which physical processes control their duration and drive recovery to “normal” cyclic activity*”. The results of our work suggest that grand minima and maxima may simply be the result of interference that occurs between the four decadal components of SSN. Usoskin et al (2021) examined, specifically, the reconstructed SSN periodicity during the Sporer Minimum. A spectral analysis (their Figure 15) showed that octal periodicity was dominant during the Sporer minimum, roughly 1400 to 1600. This contrasts with the average, 11.6 years, of their estimated cycle lengths during the Sporer Minimum, listed in their Table 1. The contradiction can be partly resolved by (a), the concept, proposed here, of the four decadal cycles entering intervals of destructive interference and being exceeded during these intervals by octal and/or decadal-plus cycles during grand minima, and (b), the Usoskin et al (2021) analysis recording two octal, (~8 year), cycles as one decadal-plus, (~15 year), cycle during 1400 – 1418, 1436 – 1453 and 1478 – 1494. Usoskin et al (2021) conclude that “*grand minima form a separate mode of solar* *activity with solar activity dropping below the threshold of sunspot formation during their deepest phases*". This contrasts with the concept proposed here of sunspots being generated at four closely spaced near-decadal periods with the overall number of sunspots transitioning between grand minima and maxima, at centennial scale, due to destructive and constructive interference of the decadal components and, during intervals of destructive decadal interference, of sunspot generation being primarily of octal or decadal-plus periodicity.

Figure 9 below shows a wavelet analysis of the Usoskin et al (2021) SSN reconstruction 971 to 1899. During the Sporer Minimum, ~1400 to ~1600, this analysis clearly shows the rapid decrease in decadal periodicity at around 1400, leading into a prolonged grand minimum of decadal periodicity that extends to 1600. The proposal here is that during the 1400 – 1600 interval the four decadal components interfere destructively to near zero and decadal periodicity is largely replaced by a combination of mainly octal as well as, weaker, decadal-plus components. The rapid falloff in the decadal component at around 1400 is consistent with the shah model scenario of Figure 7A depicting grand minima generation. The spectral content of solar activity during the Sporer Minimum, (Usoskin et al 2021, Dash et al 2023), is consistent with our assessment. Similar rapid falloff in SSN is evident for the Oort Minimum, ~1000 to ~1100. It is suggested that during this interval weak octal and decadal-plus periodicity continues while decadal periodicity interferes to near zero. Similarly, for the Maunder Minimum, ~1600 to ~1700, decadal periodicity interferes to zero while weak octal periodicity continues.

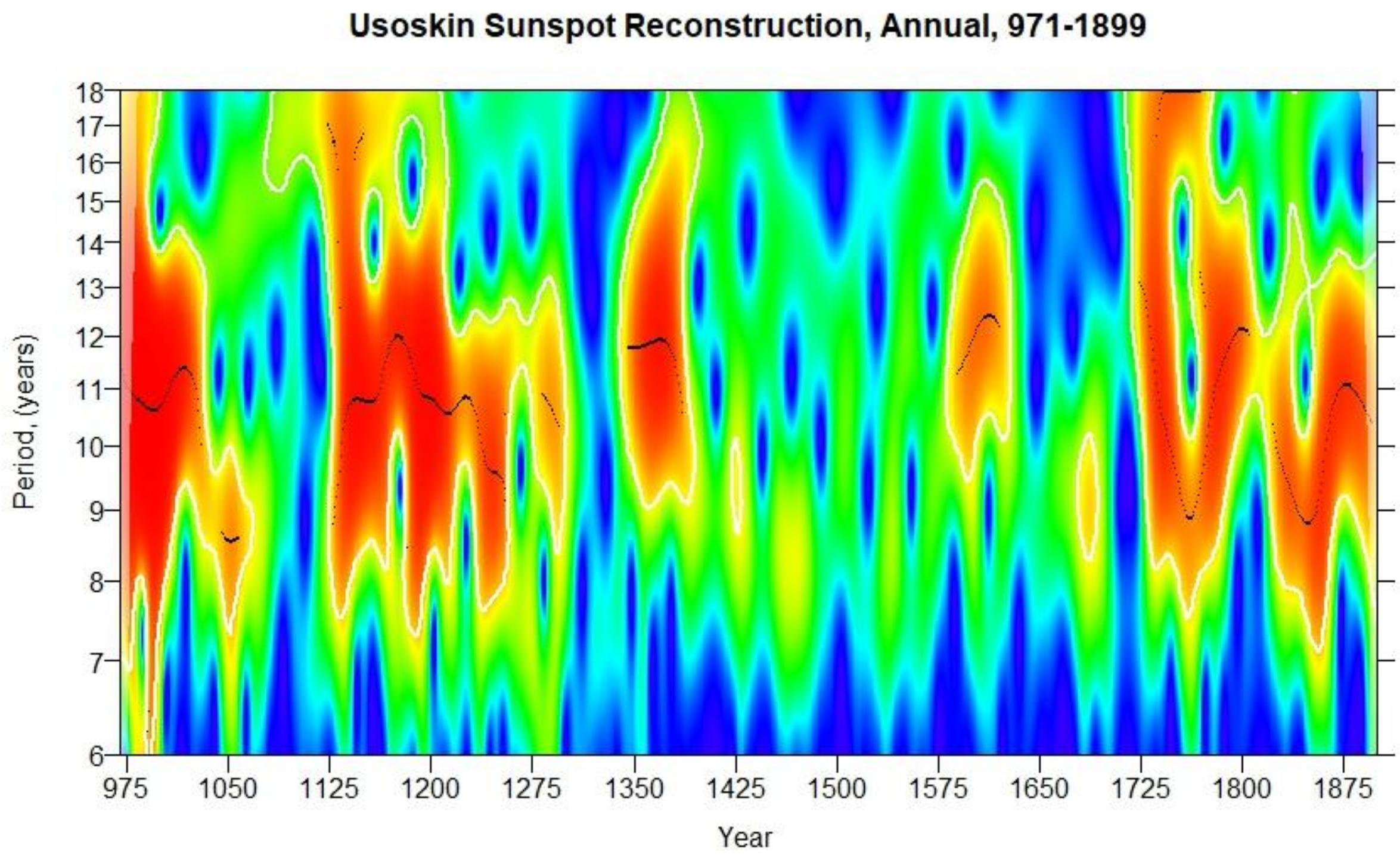

**Figure 8. A wavelet analysis of the Usoskin et al (2021) reconstructed SSN, 971 to 1899, illustrates how observed SSN variation replicates the "shah" model of long term sunspot number variation shown in Figure 7. The wavelet analysis supports the idea of components in the decadal range, average periodicity, ~11 years, transitioning slowly, at the centennial scale, between prolonged intervals of destructive interference and short intervals of constructive interference. The Sporer Grand Minimum, 1400 to 1600, replicates this behavior closely. In the absence of significant decadal periodicity the remnant SSN periodicity during grand minima is primarily octal during the Sporer and Maunder grand minima and partly octal and partly decadal-plus during the Oort grand minimum. The wavelet analysis also replicates the "shah" model of Figure 7 in that sharp falls in decadal periodicity presage prolonged grand minima, the Oort, Sporer and Maunder minima being examples. The Wolf (1300 – 1340) and Dalton (1800 – 1820) Minima replicate the converse, i.e. slow falls in decadal periodicity presage short grand minima.**

**4.3 A model of the Waldmeier Effect**. The Waldmeier Effect, Waldmeier (1935), in its different manifestations has been studied extensively, (Kane 2008, Cameron and Schussler 2008a, Karak and Choudhuri 2011, Pipin and Kosovichev 2011, Takalo and Mursula 2018, Garg et al 2019, Svalgaard and Hathaway 2020, Usoskin et al 2025, Osipova et al 2025), with no consensus on its origin. In Section 3.3 we found that interference between the summed decadal and the summed octal components of the SSN reproduced the Waldmeier Effect. Here we introduce a very simple model that supports this idea.

Figure 9A depicts the interference of two sinusoids. The stronger, nominally decadal, has a period of 10.75 years. The other, nominally octal, at one third the amplitude, has period of 8.25 years. The sum of the two sinusoids, (labeled $y_1+y_2$, solid red line), is a quasi-periodic cycle exhibiting variation in cycle amplitude and cycle length. Low amplitude cycles occur when the octal sinusoid is falling and the decadal sinusoid is rising, (dotted vertical lines, Figure 9A. High amplitude cycles occur when both sinusoids are rising, (dash-dot vertical lines, Figure 9A).

The behavior of all cycles is illustrated in Figure 9B where the octal plus decadal, $y_1+y_2$, curve is plotted, (red line), together with its time differential, (the black broken line), indicating the slope of the $y_1+y_2$ curve. Visual comparison reveals that, for all cycles, the slope of the rising part of the cycle is correlated with the peak amplitude of the cycle, a low/high initial slope correlating with a low/high peak amplitude. Note that a low/high initial slope corresponds to a high/low rise time leading to the inverse relationship of cycle rise time and cycle peak amplitude known as the Waldmeier Effect.

With this model we can also discern other "effects": A cycle with low/high final slope invariably leads to a long/short cycle length. This can be seen in the cycles around the 100 year mark in Figure 9B where the cycle lengths are labeled, 12, 11 and 10 years. The effect was called the Amplitude-Period Effect by Hathaway (2015). The correlation observed by Solanki et al (2002) between the length of cycle n-3 and the amplitude of cycle n is also evident. Another effect noticeable in Figure 9B is that the amplitudes of solar cycles n and n-3 are similar, or $A_{n-3}/A_n \sim 1$. Thus, the model supports the idea that all manifestations of the Waldmeier Effect can be explained by interference between the octal and decadal components of the SSN. Adding a decadal-plus cycle at period 14 years and amplitude one quarter the decadal amplitude does not significantly alter the model and findings.

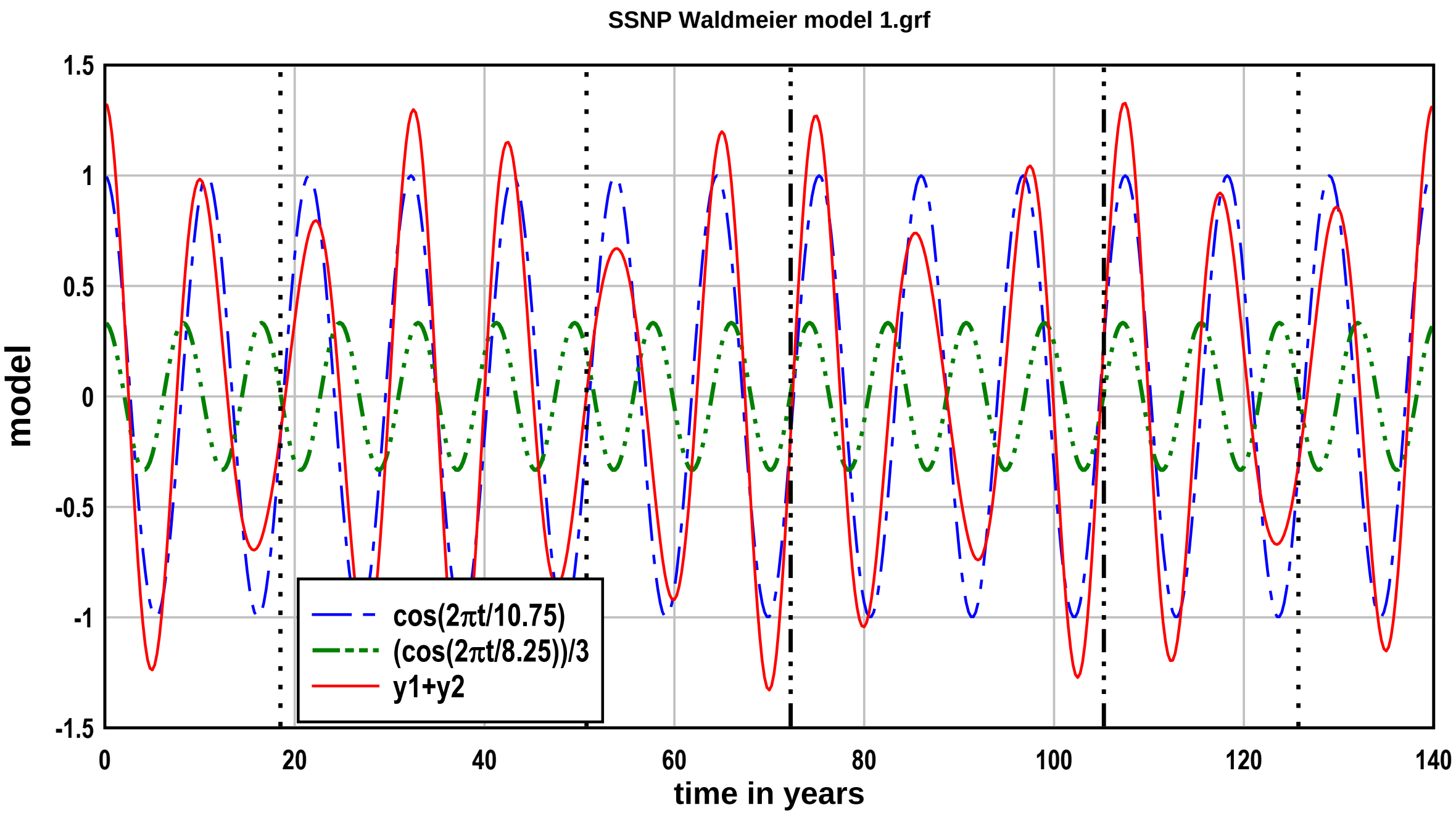

(A)

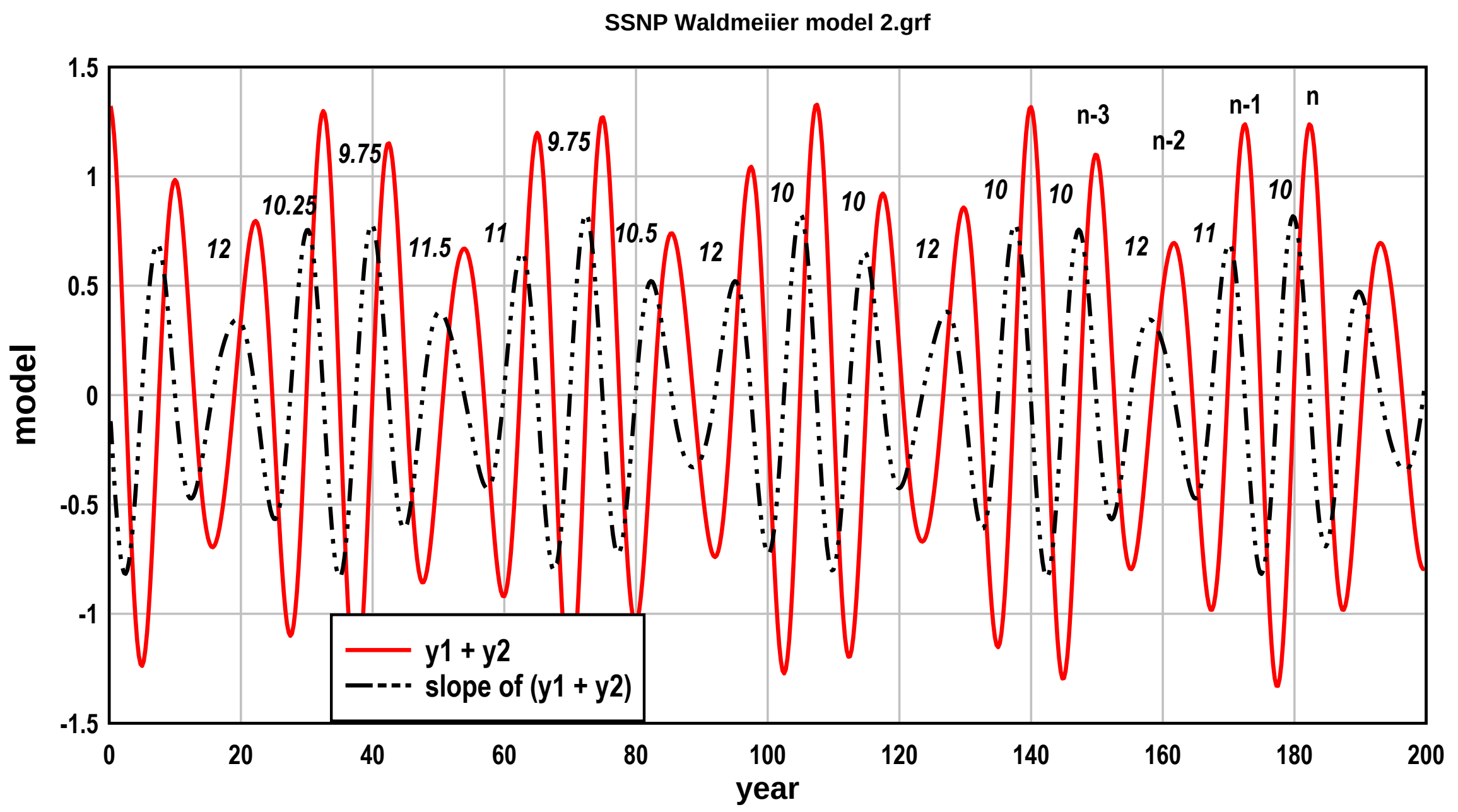

(B)

**Figure 9. (A) A model of the Waldmeier Effect based on the interference of a decadal component and an octal component, amplitude about one third the decadal amplitude. The sum of the two components, y1+y2, is the full red line. The dotted reference lines mark times when the decadal component is rising at the same time the octal component is falling. This leads to a long rise time, T, of the sum which reaches its peak, $SSN_{max}$, when the decadal and octal components are in anti-phase giving a low $SSN_{max}$. Hence the Waldmeier Effect: an inverse relationship between T and $SSN_{max}$. The dash dot reference lines represent the opposite scenario – that also leads to the Waldmeier Effect. (B) The red line is the sum y1+y2 of the decadal and octal components. The broken black line is the time differential of the sum, representing the slope of the sum. Evidently, the sum and the slope of the sum are related with low/high slope (long T/short T) leading to low $SSN_{max}$/high $SSN_{max}$. Thus, the model generates this Waldmeier Effect in all cycles. Another Waldmeier Effect, peak to peak length of cycle n-3 proportional to amplitude of cycle n, Solanki et al (2002), is also evident, see the cycles marked n-3, n-2, n-1, n and note the time interval between cycle maxima labeled, e.g. *12* years.**

**4.4 On the origin of decadal-plus spectral content of SSN.** So far we have not discussed the origin of the periodicities evident in the spectrum of Figure 1B and listed in Table 1. Here we focus on the decadal-plus periods, 13.10, 13.92 and 15.05 years as measured at spectral peaks in Figure 1B and consider the possibility that the periods are associated with the conjunctions of Jupiter and Neptune, (J/N), and Jupiter and Uranus, (J/U). The average interval between the conjunctions is given by the equation $T = (1/T_J - 1/T_I)^{-1}$ where $T_J$ is the orbital period of Jupiter and $T_I$ is the orbital period of Uranus or Neptune. If the orbital periods of Jupiter, Uranus and Neptune given in the NASA planetary table https://nssdc.gsfc.nasa.gov/planetary/factsheet/planet_table_ratio.html are used the J/N conjunctions occur at intervals of 12.83 years and the J/U conjunctions occur at intervals of 13.87 years. These intervals are close to but not exactly on the 13.10 and 13.92 year periods assessed from the SILSO SSN record, Figure 1B. In the first instance one would expect planetary based periodicities to be exact. However, due to the fact that planet orbits are elliptical and at different inclinations the intervals between conjunctions can vary quite markedly when calculated more exactly using ephemeris data. Figure 10 plots J/U and J/N conjunction intervals versus time. The values were found using the ephemeris data provided online by the Institute of Applied Astronomy, https://iaaras.ru/en/dept/ephemeris/online , and determining the date of the minimum angle between the vectors from the Sun to the particular planets. Clearly, sequences of J/U conjunction intervals occur at close to 14.1 years and sequences of J/N conjunction intervals occur at close to 13.0 years. If SSN is influenced by planetary conjunctions it is expected that Fourier analysis of the SSN record would select such sequences and record higher periods than the average periods as calculated from orbital periods, i.e. SSN periodicity at 13.10 years rather than 12.83 years and SSN periodicity at 13.92 years rather than 13.87 years. Further it is evident from Figure 10 that long J/U conjunction intervals, close to 15 years, occasionally occur which could account for the small 15.05 year peak evident in the SSN spectrum, Figure1B.

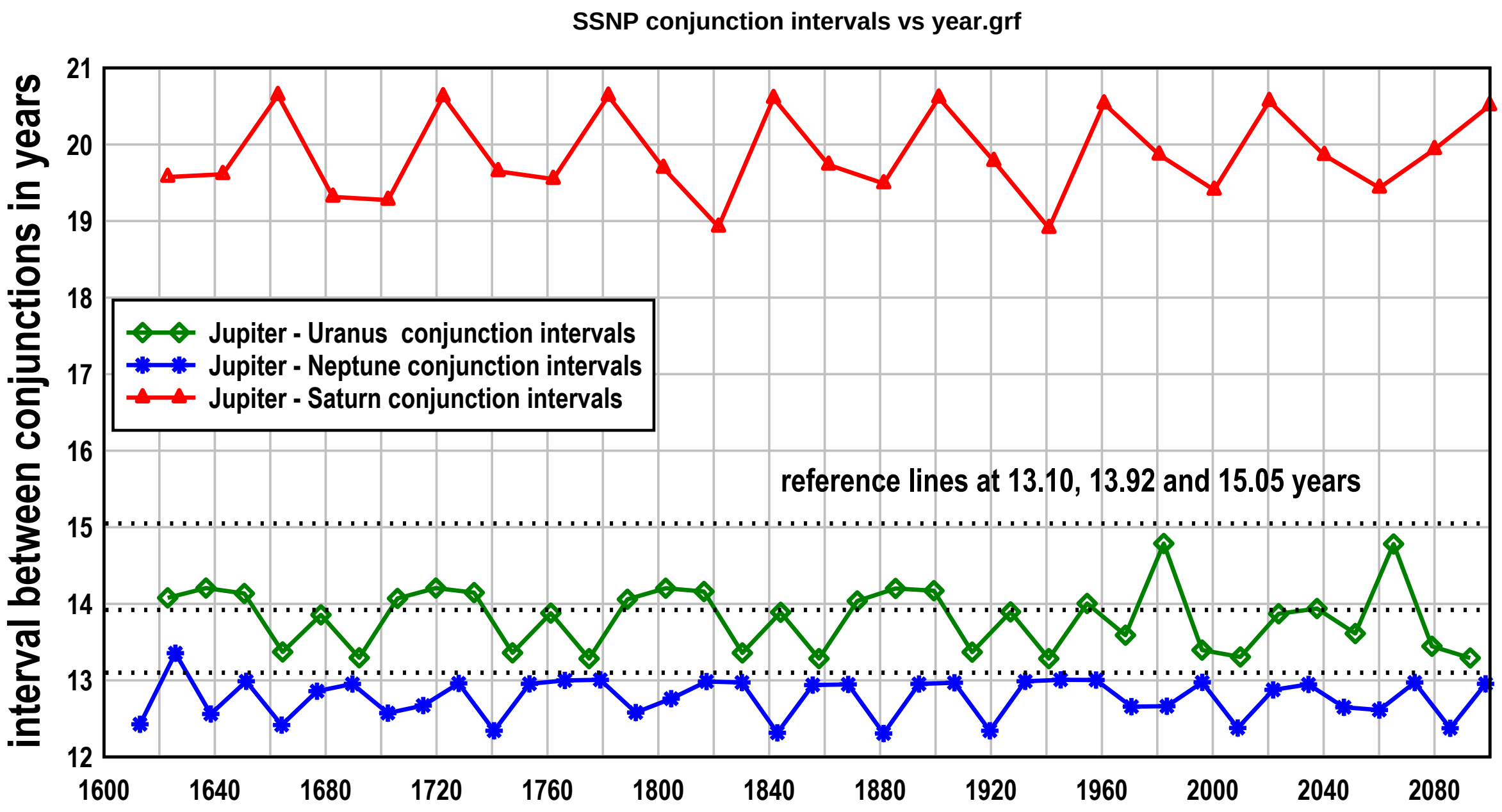

**Figure 10. Time intervals between conjunctions as estimated from ephemeris data vary significantly with time. The horizontal references lines are plotted at the periods of the decadal-plus peaks as estimated from the spectrum of SILSO SSN, Figure 1B.**

## 5. Summary and conclusion.

The analyses above are mainly based on the observed SILSO annual SSN record, 1700.5 to 2024.5 and on a reconstructed annual SSN, 971 to 1899. The main findings are:

(1) The variation in SSN anomaly is largely characterized, (r = 0.83), by three groups of periodicities, nominally the octal, decadal and decadal-plus groups, in the period range 8 to 15 years.

(2) The SSN variation due to the four components in the decadal group of periodicities, 10 to 12 years, dominates the SSN variation but is occasionally exceeded in amplitude by the octal group or the decadal-plus group.

(3) When the eleven components corresponding to the periodicities in the three groups are found by the INF method, fitted to sinusoids at solar cycle 24, projected forward and summed, the sum forecasts a solar grand minimum in SSN extending from about 2030 to 2150.

(4) The forecast was validated by a back projection of sinusoids, fitted to components at solar cycle 24, that reproduced the SSN variation in the Dalton Minimum and the Maunder Grand Minimum.

(5) A model, the “shah” model, connecting the generation of grand minima/maxima to interference between the components in the decadal group was developed and was shown to reproduce grand minima/maxima the features of which were consistent with features obtained from wavelet analysis of reconstructed SSN.

(6) Interference between the summed octal and the summed decadal components of SSN reproduced the Waldmeier Effect in its various manifestations.

(7) An analysis that isolated four long period components of a reconstructed SSN record, 971 to 1899, and projected the components forward correctly forecast the Gleissberg Minimum and the Modern Maximum and forecast a grand minimum between 2000 and 2150.

The forecasts of the next grand minimum obtained in this work depend on the persistence of cycles identified in the SILSO SSN record and in a reconstructed SSN record. Evidence supporting the persistence of cycles included successful back projection from the SILSO SSN record and successful forward projection from the reconstructed 971 to 1899 record. The findings herein are consistent with recent observational and simulation studies of solar magnetic behavior during grand minimum events, (Saha et al 2022, Inceoglu 2024), that confirms the persistence of octal and decadal-plus cycles during grand minima. Finally, based on this forecast, the next several decades should be ideal for manned space exploration.

**Acknowledgments:** The authors appreciate the assistance of Dr Donald Meyers in editing this paper.

**Funding:** This research did not receive any specific funding.

**Conflicts of interest:** The authors declare no Conflicts of Interest.

**Data availability statement:** The data records analyzed in this paper are:

(1) The SILSO annual SSN record, 1700.5 - 2024.5, downloaded from https://www.sidc.be/SILSO/datafiles, Source: WDC-SILSO, Royal Observatory of Belgium, Brussels, DOI: https://doi.org/10.24414/qnza-ac80
and

(2) The reconstructed annual SSN record, 971 – 1899, Usoskin et al. (2021), downloaded from https://cdsarc.cds.unistra.fr/ftp/J/A+A/649/A141/osn.dat